\documentclass[review]{elsarticle}
\usepackage{times}
\usepackage{amsfonts}
\usepackage{amsmath}
\usepackage{amssymb}
\usepackage{natbib}
\usepackage{graphicx}
\usepackage{enumitem}
\usepackage{xcolor}
\usepackage[colorlinks=true,linkcolor=blue]{hyperref}
\usepackage[outdir=./]{epstopdf}
\usepackage[normalem]{ulem}
\usepackage{longtable}
\usepackage{lscape}
\usepackage{supertabular}

\def\aap{A\&A}
\def\aj{AJ}
\def\apj{ApJ}
\def\apjl{ApJL}
\def\apjs{ApJS}

\def\jcap{JCAP}
\def\mnras{MNRAS}
\def\na{New Astronomy}

\def\nat{Nature}
\def\prd{Physical Review D}

\usepackage{lineno,hyperref}
\modulolinenumbers[5]

\journal{Journal of \LaTeX\ Templates}

\begin{document}

\begin{frontmatter}

\title{
SPARC HSBs, and LSBs, the surface density of dark matter haloes, and MOND.
}
%

\author{Antonino Del Popolo\fnref{myfootnote}}
\address{}


\cortext[mycorrespondingauthor]{Antonino Del Popolo}
\ead{antonino.delpopolo@unict.it}

\address[mymainaddress]{Dipartimento di Fisica e Astronomia, University Of Catania, Viale Andrea Doria 6, 95125 Catania}
\address[mysecondaryaddress]{Institute of Astronomy, Russian Academy of Sciences, 119017, Pyatnitskaya str., 48 , Moscow, Russia}

\begin{abstract}
In this paper, we use SPARC's HSBs, and LSBs galaxies to verify two issues.
The first one is related to one claim of \citep{Donato} D09, namely: is the DM surface density (DMsd) a constant universal quantity, equal to $\log{(\rm \Sigma/M_\odot pc^{-2})}=2.15 \pm 0.2$, or does it depend on the baryon surface density of the system? The second one, is based on a MOND prediction that for HSBs the DMsd is constant, and equal to $\log{(\rm \Sigma/M_\odot pc^{-2})}=2.14$, while for LSBs the surface density is not constant and takes 
values that are smaller than for HSBs and D09 prediction \citep{Milgrom2009}.
We find that HSBs shows a constant DMsd vs magnitude as in D09, and a constant DMsd vs $\Sigma_{\rm eff}$  as in MOND prediction, for HSBs with $\Sigma_{\rm eff}>200 L_\odot/pc^2$, and $\Sigma_{\rm eff}>300 L_\odot/pc^2$. However, the value of the DMsd is larger, $\Sigma \simeq 2.61$ (in the case of the DMsd-magnitude with  
$\Sigma_{\rm eff}>300 L_\odot/pc^2$), and $\Sigma \simeq 2.54$ (in the case of the surface DMsd-surface brightness with 
$\Sigma_{\rm eff}>200 L_\odot/pc^2$). This value slightly depends on the threshold to determine wheter a galaxy is HSB.  
In the case of LSBs, for $\Sigma_{\rm eff}<100 L_\odot/pc^2$, and $\Sigma_{\rm eff}<25 L_\odot/pc^2$,  the surface density vs magnitude, for lower magnitudes, is approximately equal to that predicted by D09, but several galaxies, for magnitude $M>-17$ have smaller values than those predicted by D09. The DMsd vs $\Sigma_{\rm eff}$ shows a similar behavior in qualitative, but not quantitative, agreement with MOND predictions. In summary, in the case of HSBs both D09 and MOND are in qualitative, but not quantitative, agreement with the data. In the case of LSBs D09 is mainly in disagreement with the data, and MOND only in
qualitative agreement with them.
\end{abstract}

\begin{keyword}
\texttt{Galaxies; Alternative theory of gravity; galaxies surface density}
\end{keyword}

\end{frontmatter}

%
%

\section{Introduction}

In spite of the fact that the $\Lambda$CDM model, is a very good model in describing the observations on 
intermediate scales \citep{Spergel,Kowalski,Percival,2011ApJS..192...18K,DelPopolo2007,2013AIPC15482D,2014IJMPD2330005D}, and cosmology scales \cite{2011ApJS..192...18K,DelPopolo2007,2013AIPC15482D}, it shows several drawbacks. To start with we recall the cosmological constant problem \citep{Weinberg, Astashenok}, and the cosmic coincidence problem \citep{Velten2014}. Apart the mentioned problems several others are present. 
At large scale there is the unsolved tensions between the {the current value of the Hubble parameter}, $H_0$, having different values when measured with different methods, for example using the CMB, and 
supernovae and stars in the relatively recent universe \citep{DiValentino2021}. 
Moreover, there is another tension between the data of Planck 2015 and the 
$\sigma_8$ {growth rate of perturbations} \citep{maca}, and with CFHTLenS weak lensing \citep{raveri} data.
In the large-angle fluctuations in the CMB are present a power hemispherical asymmetry \citep{eriksen,hansen,jaf,hot,planck1,akrami}, a quadrupole-octupole alignment \citep{schwa,copi,copi1,copi2,copi3}, and a cold spot \citep{cruz,cruz1,cruz2}. 

Other problems are present on smaller scales ($\simeq 1-10$ kpcs). Those often recalled are 
the "Cusp/Core" problem, namely the discrepancy between the cuspy profiles obtained in dark matter (DM) only N-body simulations (\citep{nfw1996,nfw1997,Navarro2010}, and the 
observations of dwarf spirals, dwarf spheroidals (dSphs), and Low Surface Brightness (LSBs) galaxies showing cored profiles (\citep{Moore1994,Flores1994,Burkert1995,deBloketal2003,Swatersetal2003,DelPopolo2009,DelPopoloKroupa2009,2012MNRAS419971D,deBloketal2003,Swatersetal2003,DelPopolo2009,DelPopoloKroupa2009,DelPopoloHiotelis2014})
. The ``missing satellite problem" consists in the discrepancy between the number of satellites or subhaloes predicted by DM only N-body simulations (\citep{Klypin1999,Moore1999}), and the number of subhaloes really observed. Moreover, the subhaloes obtained in simulations are too dense with respect to those observed around the Milky Way \citep{GarrisonKimmel2013,GarrisonKimmel2014}.  The last is dubbed ``too-big-to-fail" problem. Finally, we cite the issue of the location on planes of satellite galaxies of the Milky Way and M31 \citep{pawl}, difficult to explain. This is dubbed ``satellites planes problem", but recently \citep{Sawala2022} showed a solution. The quoted problems have been attacked from different fronts. Some authors proposed to modify the power spectrum (e.g. \citep{2003ApJ59849Z}), others proposed to modify either the particles constituting DM (\citep{2000ApJ542622C,2001ApJ551608S,2000NewA5103G,2000ApJ534L127P}), or the theory of gravity (\citep{1970MNRAS1501B,1980PhLB9199S,1983ApJ270365M,1983ApJ270371M,Ferraro2012}). Apart these drastic solutions, other astrophysical solutions have been proposed. They are based on the role of baryons in ``heating" DM. A well known mechanism is that related to supernovae feedback (\citep{1996MNRAS283L72N,Gelato1999,Read2005,Mashchenko2006,Governato2010}),
and another one is related to the transfer of energy and angular momentum from baryons to DM through dynamical friction (\citep{El-Zant2001,El-Zant2004,2008ApJ685L105R,DelPopolo2009,Cole2011,2012MNRAS419971D,Saburova2014}. 

In this context, scaling relations are very helpful to understand complex phenomena.
\cite{Kormendy2004}, by fitting the rotation curves of 55 galaxies with a pseudo-isothermal profile,
obtained some relations among the DM halos parameters.

They introduced the quantity $\Sigma=\mu_{0D} =\rho_0 r_0$\footnote{$r_0$ is the core radius of the pseudo-isothermal profile, and $\rho_0$ its central density}, which behaves as a DMsd. 
According to their studies, for late type galaxies, $\Sigma=\mu_{0D} \simeq 100 M_{\odot}/pc^{2}$ independently from galaxy luminosity. This result has been studied and verified by several authors. \cite{Donato} (D09), in agreement with \cite{Kormendy2004}, found again a quasi-universality of $\Sigma$, by analysing a set of 1000 galaxies (spirals, dwarfs, ellipticals, etc). 
Promptly, \cite{Milgrom2009}, showed that in MOND $\Sigma$, in the Newtonian regime, has a similar behavior to that described by D09, but noticed that in the case of galaxies having a low surface density $\Sigma$ has a smaller value. Recently, \citep{DelPopolo2023} extended the work of \cite{Milgrom2009} to spiral galaxies showing that for high baryon DMsd $\Sigma$ is constant in agreement with \cite{Milgrom2009}, and D09, and for low baryon surface density $\Sigma$ decreases. 
%
%

%
%

From the literature is known that the Burkert profile gives good fits to LSBs, and dwarfs rotation curves (\citep{Gentile2004,Gentile2007,DelPopolo2009}. In the case of 
giant galaxies, and ellipticals the previous conclusion is not valid (\citep{Simon2005,THINGS,2012MNRAS42438D,Simon2005}).

%
%

Several authors obtained results in disagreement with those of D09.  

By means of a much larger sample of that of D09 \cite{Boyarsky} showed that dark matter column density systematically increases with the mass of the halo. 
%
%
\cite{CardoneTortora2010} showed that the column density, and the Newtonian acceleration,  correlates with different quantities
in agreement with \cite{Boyarsky}, and in disagreement with D09 results.

%
%

In early-type galaxies, \cite{Napolitano2010} did not find the existence of a universal DMsd. 

\cite{2013MNRAS4291080D}, in agreement with \cite{Napolitano2010}, \cite{CardoneTortora2010}, and \cite{Boyarsky}, found signs of a correlation of DMsd with $M_{200}$. 

\cite{CardoneDelPopolo2012} found a a correlation between the Newtonian acceleration and the virial mass $M_{\rm vir}$. 

Several correlations between the surface density and a series of quantities are shown by 
\citep{Saburova2014}, similarly to what found by \citep{Zhou2020}. In the last paper the DMsd was obtained by fitting the rotation curves of the galaxies with a Burkert profile, and using the {Markov Chain Monte Carlo} (MCMC) method to infer the values of the parameters.

In this paper, using SPARC's HSBs, and LSBs we want to study the behavior of the DMsd, to see whether it is either constant as claimed by D09, or it depends on the baryon surface density according to MOND. To this aim, we study the DMsd vs the magnitude and the baryon surface density in HSBs, and LSBs. HSBs shows a constant behavior of the DM, while LSBs shows that the behavior depends from the baryon surface density.

The paper is organized as follows. In Sect. 2, we introduce the SPARC sample and the analysis that was performed in a previous paper (\citep{Zhou2020}) to obtain the DMsd. In Sect. 3, we discuss the results, and Sect. 4 is devoted to discussion.

\section{SPARC data set, and data analysis: a summary}

In this section, we give a summary of the SPARC data used, and the analysis on the data, as performed by \citep{Zhou2020}. SPARC ($\it Spitzer$ Photometry and Accurate Rotation Curves) dataset \citep{Lelli2016} is constituted by 175 late-type galaxies with high quality rotation curves obtained from $\rm HI/H\alpha$ studies, and with surface photometry at 3.6 $\mu \rm m$, giving the mass-to-light ratio $\Upsilon_*$ conversion factor. The gas mass is provided by 21 cm observations. Almost all SPARC galaxies have a disc structure, with some having also bulges. The baryon component is constituted by the disc, bulge, and gas components. The morphologies present in SPARC goes from SO to Irr. \citep{Zhou2020} obtained the DM surface density by fitting the rotation curves with a Burkert profile
\begin{equation}
\rho(r)=\frac{\rho_0 r_0^3}{(r+r_0)(r^2+r_0^2)},
\end{equation}
where $r_0$, and $\rho_0$ represent the scale radius and the central density of the halo, respectively. 

The total rotational velocity is given by
\begin{eqnarray}
V_{\rm tot}^2=V_{\rm DM}^2+V_{\rm bar}^2=V_{\rm DM}^2+\Upsilon_{\rm d}V^2_{\rm disc}+\Upsilon_{\rm b}V^2_{\rm bulge}+V^2_{\rm gas},
\end{eqnarray}
where $V_{\rm disc},~V_{\rm bulge},~V_{\rm gas}$ {are the velocities of the baryonic component}, and 
$V_{\rm DM}$ {is that of the dark matter component}, given by 
\begin{eqnarray}
\frac{V_{\rm DM}^2}{V_{200}^2}=\frac{C_{200}}{x}\frac{\ln{(1+x)}+\frac{1}{2}\ln{(1+x^2)}-\arctan{x}}{\ln{(1+C_{200})}+\frac{1}{2}\ln{(1+C_{200}^2)}-\arctan{C_{200}}},
\end{eqnarray}
and the concentration $C_{200}$ and the rotation velocity $V_{200}$ at the virial radius 
$r_{200}$ are given by
\begin{eqnarray}
\label{eq1}
C_{200}=r_{200}/r_0,~~V_{200}=10C_{200}r_0H_0,
\end{eqnarray}
where $H_0$ is the Hubble constant (chosen to be $73~\mathrm{Km\,s^{-1}\,Mpc^{-1}}$).
The mass-to-light ratios for the disc and bulge component, are given by 
$\Upsilon_{\rm d}$, and $\Upsilon_{\rm b}$, respectively. 
An important point to recall is that the galaxy distance, and the disc inclination, affect the stellar components and the total observed rotational velocities, respectively. Then if the galaxy distance is changed from its value $D$ to $D'=D\delta_{\rm D}$\footnote{$\delta_{\rm D}$ is a dimensionless distance factor}, also the radius is affected and becomes $R'=R\delta_{\rm D}$, and also the baryonic component velocity which becomes $V'_{\rm k}=V_{\rm k}\sqrt{\delta_D}$\footnote{`k' denotes disc, bulge, or gas}. 
 
A change in disk inclination from $i$ to $i'=i\delta_{\rm i}$\footnote{$\delta_{\rm i}$ is a dimensionless inclination factor}, produces a change in the observed rotation curves and their uncertainties, as 
\begin{eqnarray}
V_{\mathrm{obs}}^{\prime}=V_{\mathrm{obs}} \frac{\sin (i)}{\sin \left(i^{\prime}\right)}, \quad \delta V_{\mathrm{obs}}^{\prime}=\delta V_{\mathrm{obs}} \frac{\sin (i)}{\sin \left(i^{\prime}\right)}.
\end{eqnarray} 
 
The observed rotational curve is fitted with a theoretical curve depending on several parameters:  $V_{200}$, $C_{200}$, $\Upsilon_{\rm d}$, $\Upsilon_{\rm b}$, $\delta_{\rm D}$ and $\delta_{\rm i}$. The calculation is performed through a Bayesian analysis. The posterior probability of parameter space is given by  
\begin{eqnarray}
\nonumber P(V_{200},C_{200},\Upsilon_{\rm d},\Upsilon_{\rm b},\delta_{\rm D},\delta_{\rm i}|\mathrm{SPARC})=\mathcal{L}(V_{200},C_{200},\\\Upsilon_{\rm d},\Upsilon_{\rm b},\delta_{\rm D},\delta_{\rm i}|\mathrm{SPARC})
P(V_{200},C_{200},\Upsilon_{\rm d},\Upsilon_{\rm b},\delta_{\rm D},\delta_{\rm i}),
\end{eqnarray}
where the likelihood is obtained through $\mathcal{L}\sim e^{-\chi^2/2}$, and where $\chi^2$ is given by 
\begin{eqnarray}\label{eq:chi2}
\chi^2=\sum_{k=1}^N \left(\frac{V_{\rm tot}(R'_{\rm  k};V_{200},C_{200},\Upsilon_{\rm d},\Upsilon_{\rm b},\delta_{\rm D})-V'_{\rm obs,k}}{\delta V_{\rm obs,k}^{'}}\right)^2,
\end{eqnarray}
In the previous relation, $N$ is the number of data point for each galaxy, 
$V'_{\rm obs,k}$ and $\delta V'_{\rm obs,k}$ are the observed rotation curve and its uncertainty at the radius $R_{\rm k}$. The total rotation velocity $V_{tot}$ at the radius $R'_k$ depends on galactic parameters $\{\Upsilon_{\rm d},\Upsilon_{\rm b},\delta_{\rm D}\}$, and 
halo parameters $\{V_{200},C_{200}\}$. {Since prior probabilities of parameters are uncorrelated it is given by}
\begin{equation}
P(V_{200},C_{200},\Upsilon_{\rm d},\Upsilon_{\rm b},\delta_{\rm D},\delta_{\rm i})=P(V_{200})P(C_{200})P(\Upsilon_{\rm d})P(\Upsilon_{\rm b})P(\delta_{\rm D})P(\delta_{\rm i}). 
\end{equation}

The priors on galactic parameters are set as in \citep{Li:2018rnd}: 
on $\delta_{\rm D}$ and $\delta_{\rm i}$ are imposed Gaussian priors around 1 with standard deviations given by the observational relative errors. On $\Upsilon_*$ is used a log-normal prior  around their fiducial values $\Upsilon_{\rm d}=0.5$ and $\Upsilon_{\rm b}=0.7$ with a standard deviation of 0.1 dex. In the case of the halo parameters is used a flat prior with $10<V_{200}<500~\mathrm{km\,s^{-1}},1<C_{200}<100$. 
Maximizing the posterior probability, one gets the best fitting value. The values of $\rho_0 r_0$, obtained with the previous method by \citep{Zhou2020}, are in the second column of Table 1, while the values of the luminosity, and the effective surface brightness (taken from SPARC's webpage \verb http://astroweb.cwru.edu/SPARC/SPARC_Lelli2016c.mrt) can be found in column 3, and 4 of Table 1, respectively.

%
%

{As already reported, by means of the Bayesian method, \citep{Zhou2020} fitted 175 galaxy rotation curve of the SPARC sample. The fit to three representative galaxies has been shown in Fig. 1 of \citep{Zhou2020}. Burkert's profile gives a good fit to the galaxies studied. The galaxies studied, except five, have a reduced $\chi^2<10$, and the best fitting values and the reduced $\chi^2$ for the SPARC sample are listed in Table 1 of \citep{Zhou2020}.
}

\section{Results}

\begin{figure*}[!ht]
 \centering
\includegraphics[width=15cm,angle=0]{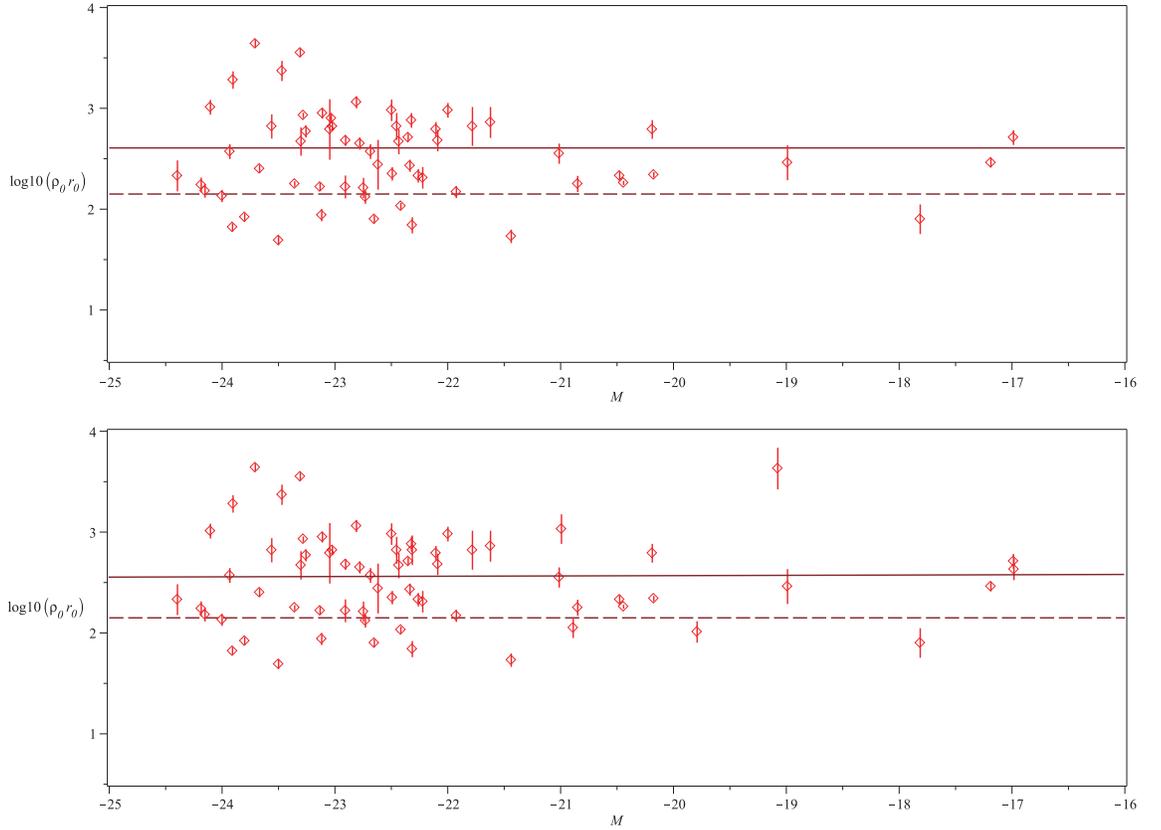} 
 \caption[justified]{Upper panel: the DMsd in terms of the magnitude, $M$ for HSBs. The data points with error-bars represent the HSBs with surface brightness larger than 300 $ \rm L_\odot/pc^2 $. The dashed line is the D09 value, and the solid line the fit to the data. Lower panel: same as the upper panel but for HSBs with surface brightness larger than 200 $ \rm L_\odot/pc^2$. 
}
 \label{fig:comparison}
\end{figure*}

\begin{figure*}[!ht]
 \centering
\includegraphics[width=15cm,angle=0]{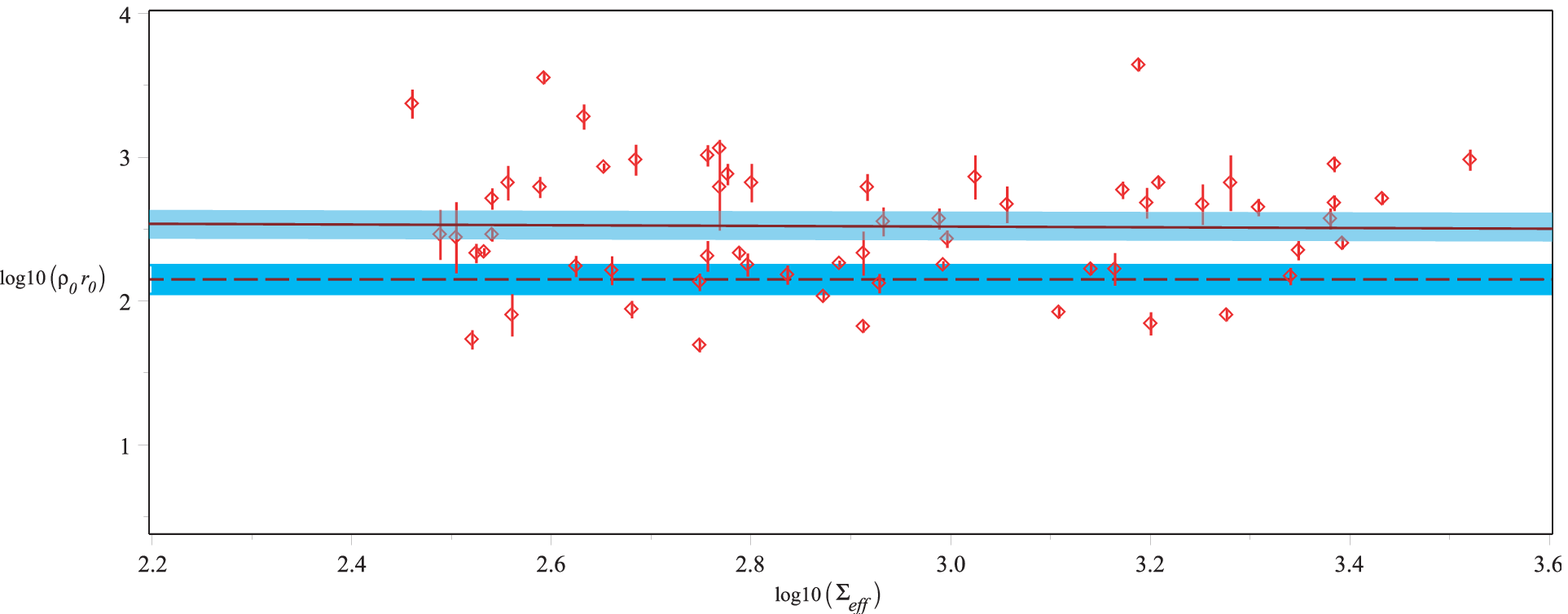} 
\includegraphics[width=15cm,angle=0]{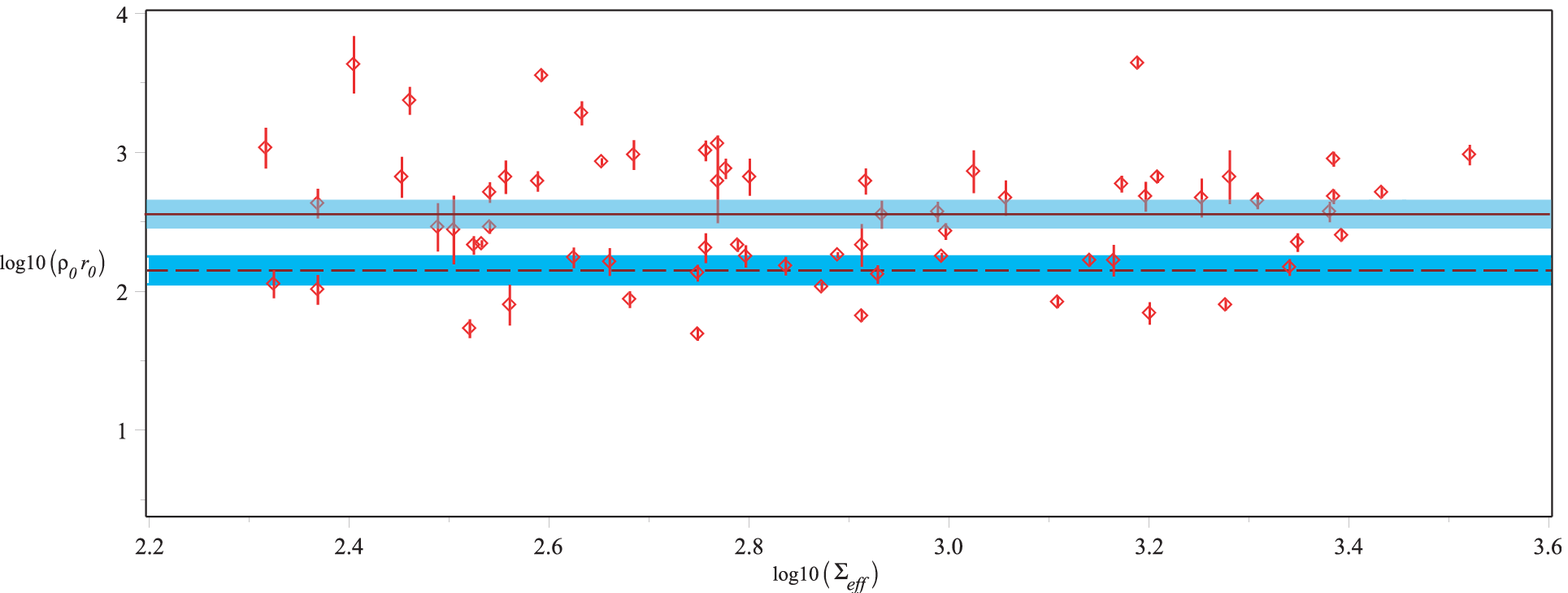}
 \caption[justified]{Upper panel: the DMsd in terms of the effective surface brightness, $\Sigma_{\rm eff}$ for HSBs. The data points with error-bars represent the HSBs with surface brightness larger than 300 $ \rm L_\odot/pc^2 $. The dashed line is the D09 value, the shaded region the $1 \sigma$ confidence level region obtained using MOND, and the solid line the fit to the data. Lower panel: same as the upper panel but for HSBs with surface brightness larger than 200 $ \rm L_\odot/pc^2$.
}
 \label{fig:comparison}
\end{figure*}

\begin{figure*}[!ht]
 \centering
\includegraphics[width=15cm,angle=0]{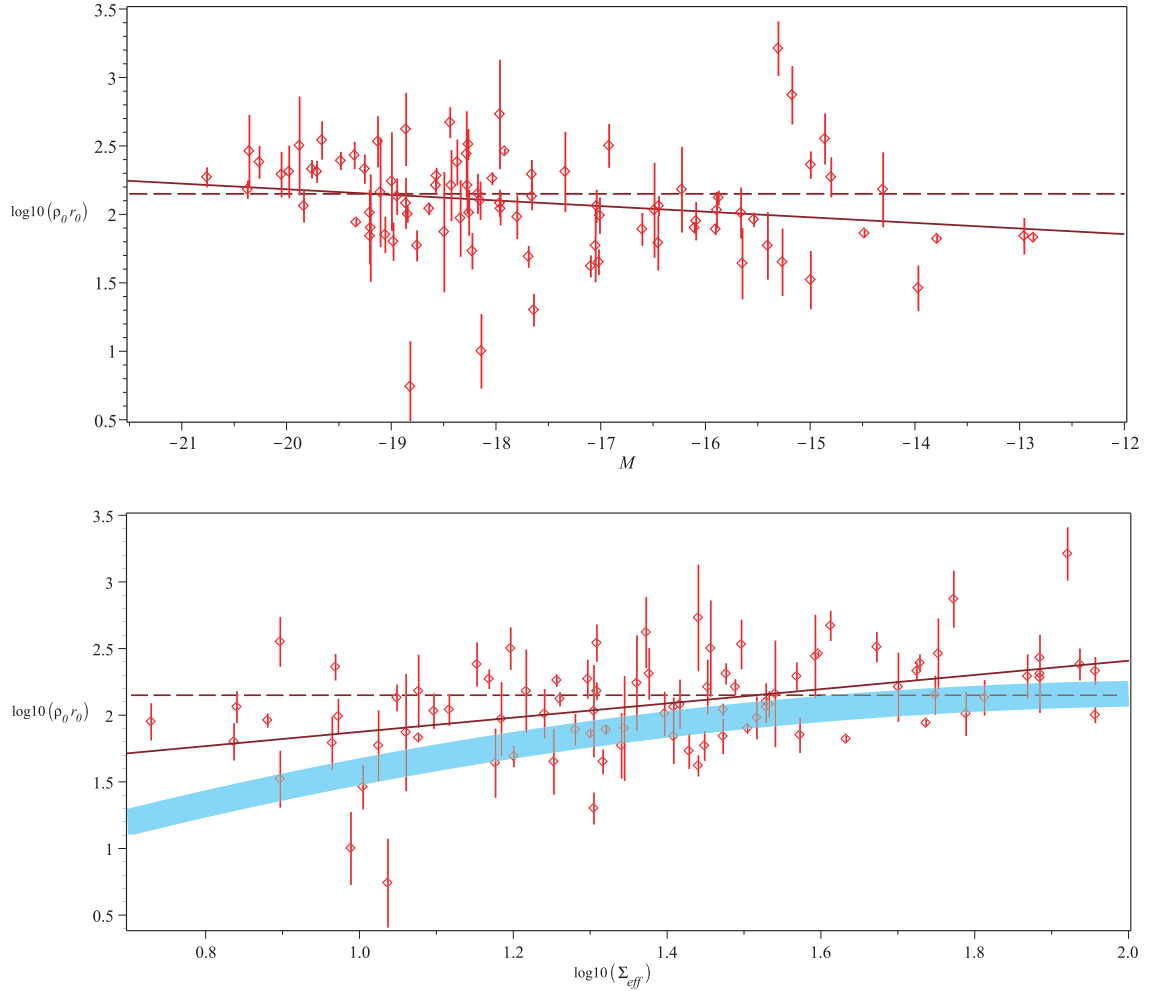}
 \caption[justified]{Upper panel: DMsd in terms of the magnitude, $M$ for LSBs. The data points with error-bars represent the LSBs with surface brightness smaller than 100 $ \rm L_\odot/pc^2 $. The dashed line is the D09 value, and the solid line the fit to the data. Bottom panel: the DMsd in terms of the effective surface brightness, $\Sigma_{\rm eff}$ for LSBs with surface brightness smaller than 100 $ \rm L_\odot/pc^2 $. The dashed line is the D09 value, and the solid line the fit to the data. The shaded region the $1 \sigma$ confidence level region obtained using MOND.
}
 \label{fig:comparison}
\end{figure*}

\begin{figure*}[!ht]
 \centering
\includegraphics[width=15cm,angle=0]{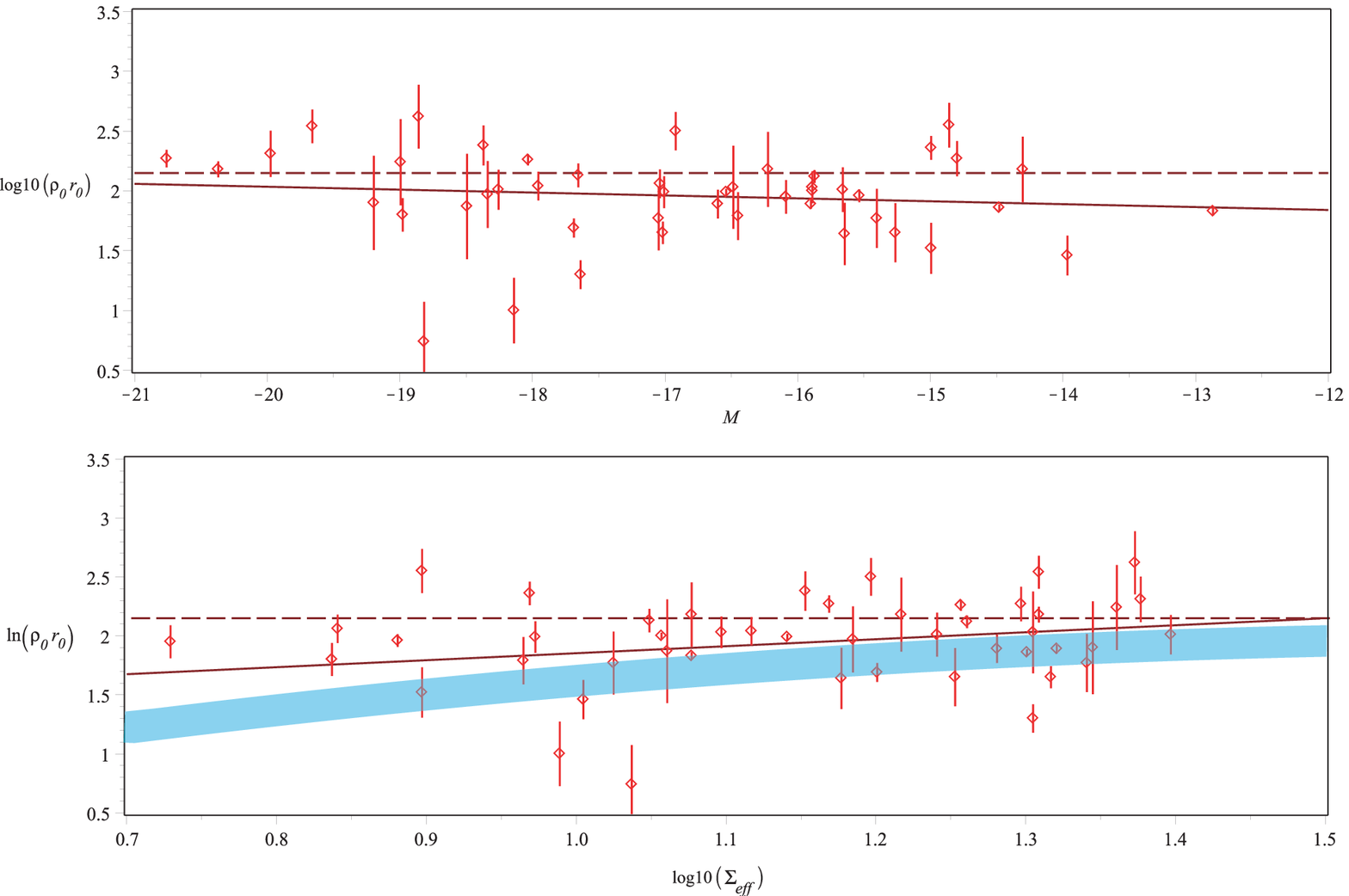} 
 \caption[justified]{Same as Fig. 3 but for DMsd smaller than 25 $ \rm L_\odot/pc^2 $.
}
 \label{fig:comparison}
\end{figure*}

As discussed in the abstract, in this paper we have two goals. The first one is related to one claim of D09, namely that the DMsd is a constant universal quantity, equal to 
$\log{(\rm \Sigma/M_\odot pc^{-2})}=2.15 \pm 0.2$. The second one, a MOND prediction, namely that for HSBs the DMsd is constant, and equal to $\log{(\rm \Sigma/M_\odot pc^{-2})}=2.14$, while for LSBs the DMsd is not constant and has smaller values than for HSBs. At the first order the DMsd behaves as $\Sigma \simeq \sqrt{\Sigma_b \Sigma_M}$. In order to answer to these two questions, we used the SPARC ($\it Spitzer$ Photometry and Accurate Rotation Curves) sample \citep{Lelli2016}, constituted by 175 late-type galaxies with high quality rotation curves. We divided the sample in HSBs, and LSBs. In order to distinguish between these two classes of objects, we recall that the division between LSBs and HSBs is at 23 $\rm mag/arcsec^2$ in B band. Recalling that in SPARC we have a photometry at 3.6 $\rm \mu m$, an estimate of the threshold between LSBs and HSBs can be reasonably put at the effective surface brightness $\Sigma_{\rm eff}=200 L_\odot/pc^2$ (or 100 $M_\odot/pc^2$ for $M/L=0.5$)\footnote{Notice that the surface brightness distribution in SPARC is continuous even if the definition has a certain degree of arbitrariness}. 

In the paper, we consider several thresholds. For the HSBs, we use values of the effective surface brightness, $\Sigma_{\rm eff}$, $>200 L_\odot/pc^2$, and $>300 L_\odot/pc^2$. For LSBs values of $\Sigma_{\rm eff}<100 L_\odot/pc^2$, and $\Sigma_{\rm eff}<25 L_\odot/pc^2$. 
For a given value of $M/L$, the quoted threshold corresponds to thresholds in mass surface density 
$\rm M_\odot/pc^2$. The results are plotted in Figs. 1-4. 
In Fig. 1, we plot the DMsd in terms of the magnitude, $M$. The DMsd of the HSBs is represented by the data points with error-bars. The data points, surface density of galaxies and errors, are obtained by 
\citep{Zhou2020} (Table A. 1). In the upper panel we used the galaxies (HSBs) with surface brightness $> 300 \rm L_\odot/pc^2 $. The lower panel represents the same quantities as the upper panel but in this case for HSBs with surface brightness larger than 200 $ \rm L_\odot/pc^2$. 
In both panels, the dashed line is the D09 value, and the solid line the fit to the data. 
Both plots give the same information: the DMsd is constant, independent on the magnitude. This is qualitatively in agreement with D09, but the value of the constant DMsd is not $\log{(\rm \Sigma/M_\odot pc^{-2})}=2.15 \pm 0.2$, but larger ($\log{(\rm \Sigma/M_\odot pc^{-2})}=2.61$, for the upper panel). 
Another difference between D09 result and ours is that D09 claims that their result is valid for all kind of galaxies: LSBs, and HSBs. In reality, their conclusions is based on the analysis of dwarf spheroidal satellites of the Milky Way. They include only one well studied LSB: NGC 3741. In other terms their claim that their result can be applied to all galaxies is not correct. The D09 result and our are in qualitative agreement because they are mainly using HSBs as we are doing in this first part of the analysis. Then HSBs are characterized by a constant DMsd. 
In Fig. 2, we plot the surface density in terms of the effective surface brightness, $\Sigma_{\rm eff}$. As in Fig. 1, the data points with error-bars represent the HSBs with surface brightness larger than 300 $ \rm L_\odot/pc^2 $ (upper panel), and 200 $ \rm L_\odot/pc^2 $ (lower panel). The dashed line is the D09 value, and the solid line the fit to the data. We also plotted, as a shaded region, the $1 \sigma$ confidence level region obtained using MOND (see Appendix). The plot gives two information: a) the DMsd is constant and does not depend on $\Sigma_{\rm eff}$. b) MOND predicts a constant value of the DMsd equal to $\log{(\rm \Sigma/M_\odot pc^{-2})}=2.14$ almost identical to that of D09, qualitatively in agreement with our result, for what concerns the constancy, but in disagreement with the value that we obtain, $\log{(\rm \Sigma/M_\odot pc^{-2})}=2.54$ (upper panel).  
In Fig. 3, the upper panel represents the DMsd in terms of the magnitude, $M$. The data points with error-bars represent the LSBs with surface brightness smaller than 100 $ \rm L_\odot/pc^2 $. The dashed line is the D09 value, and the solid line the fit to the data. The plot shows that the DMsd decreases with increasing magnitude (smaller luminosity). However, in the range of magnitudes $\simeq -20.5$ to $\simeq -17.5$ the surface density, within the errors, is in agreement with a flat surface density close to that of D09. For smaller magnitudes several LSBs have a smaller value than that predicted by D09. In other terms, for $\Sigma_{\rm eff} < 100  \rm L_\odot/pc^2 $, for decreasing values of the magnitude the LSBs tend to have smaller values of the surface density with respect to D09. The bottom panel in Fig. 3, represents the DMsd in terms of $\Sigma_{\rm eff}$. 
The behavior for decreasing values of $\Sigma_{\rm eff}$ is similar to that of the upper panel. 
Within the errors limits, and in the range $1.4 \preceq \log{\Sigma_{\rm eff}} \preceq 2$, the DMsd is in agreement with the D09 result, after it has smaller values than D09. 
Still in the bottom panel it is also plotted a shaded region which is the $1 \sigma$ confidence level region obtained using MOND. The plot shows that in the range $1.6 \preceq \log{\Sigma_{\rm eff}} \preceq 2$, MOND is in agreement with D09 prediction, while for smaller values of $\Sigma_{\rm eff}$ it shows a decline towards smaller values of the surface density, in agreement with the data. 
The upper panel of Fig. 4, represents the surface density in terms of the magnitude, $M$. 
The data points with error-bars represent the LSBs with surface brightness smaller than 25 $ \rm L_\odot/pc^2 $. The trend with the magnitude of the DMsd is similar to that shown in Fig. 3, even if the values of the DMsd are smaller than that of D09 since the most negative magnitudes. The bottom panel in Fig. 4, represents the DMsd in terms of $\Sigma_{\rm eff}$. 
As before the dots with error-bars represent the LSBs galaxies the shaded region is the $1 \sigma$ confidence level region obtained using MOND, the solid line the fit to the LSBs with surface brightness smaller than 25 $ \rm L_\odot/pc^2 $, and the dashed line D09 prediction. 
Within the errors limits, and in the range $1.35 \preceq \Sigma_{\rm eff} \preceq 1.5$, DMsd is in agreement with the D09 result, after it has smaller values than D09.
The behavior for decreasing values of $\Sigma_{\rm eff}$ is similar to that of the upper panel. In this case MOND predicts a surface density smaller than that of D09 in the entire range of the data. As in Fig. 3, it shows a decline towards smaller values of the DMsd, in agreement with the data. 

Summarizing the results till now reported, concerning HSBs, the DMsd is constant in terms of the magnitude (in qualitative agreement with D09) and $\Sigma_{\rm eff}$ (in qualitative agreement with MOND), but larger than the prediction of D09, and MOND. In the case of the LSBs, there is a decrease of the DMsd with the luminosity (in disagreement with D09), and a decrease of $\Sigma_{\rm eff}$ (in agreement with MOND), and it is usually smaller than the D09 prediction. MOND, is in qualitative agreement with the data predicts a decrease of the DMsd with $\Sigma_{\rm eff}$. 

\section{Discussion}  

By means of the HSBs, and LSBs in SPARC we studied the DMsd of the class of galaxies in the sample. Our interest was that of understanding whether the DM surface density is a constant universal quantity equal to $\log{(\rm \Sigma/M_\odot pc^{-2})}=2.15 \pm 0.2$, as claimed by D09 or if it assumes different classes of objects and their baryon surface density.  At the same time, we studied the MOND prediction (\citep{Milgrom2009,DelPopolo2023}), namely a constant DMsd in HSBs close to that predicted by D09, and a smaller, decreasing DMsd with decreasing surface brightness (baryon surface density) in the case of LSBs. In order to study these two issues, we grouped the SPARC sample in HSBs, and LSBs, and we also considered for these two groups of galaxies several thresholds in $\Sigma_{\rm eff}$. 
In the case of HSBs, and for galaxies characterized by $\Sigma_{\rm eff}>200 L_\odot/pc^2$, and $\Sigma_{\rm eff}>300 L_\odot/pc^2$, we found that the DMsd vs magnitude is constant, as in D09, and similarly, we found a constant DMsd vs $\Sigma_{\rm eff}$  as in MOND prediction. In spite of the fact that the result of our study, in the case of HSBs is in qualitative agreement with D09, and MOND, the value of the DMsd is larger than the one predicted by D09, and MOND. 
In the case of LSBs, we also used two thresholds: $\Sigma_{\rm eff}<100 L_\odot/pc^2$, and $\Sigma_{\rm eff}<25 L_\odot/pc^2$. 
For both thresholds, the DMsd is decreasing, with increasing value of magnitude (decreasing luminosity), in all magnitude range. In the case of $\Sigma_{\rm eff}<100 L_\odot/pc^2$, the DMsd, within the errors, and in in the range of magnitudes $\simeq -20.5$ to $\simeq -17.5$ is in agreement with a flat surface density close to that of D09. Going to smaller negative magnitudes several LSBs have a smaller value than that predicted by D09.  
In the case of LSBs with surface brightness smaller than 25 $ \rm L_\odot/pc^2 $, the trend with magnitude of the surface density is similar to that of the previous case ($\Sigma_{\rm eff}<100 L_\odot/pc^2$), even if the values of the DMsd are smaller than that of D09 since the most negative magnitudes. 
Concerning the trend of the DM surface density in terms of the $\Sigma_{\rm eff}$ for the two thresholds ($\Sigma_{\rm eff}<100 L_\odot/pc^2$, and $\Sigma_{\rm eff}<25 L_\odot/pc^2$), it is decreasing with decreasing $\Sigma_{\rm eff}$.
In the case $\Sigma_{\rm eff}<100 L_\odot/pc^2$, the DMsd is, in the errors limits, and in the range $1.4 \preceq \log{\Sigma_{\rm eff}} \preceq 2$, in agreement with the D09 result, after it has smaller values than D09. In the case $\Sigma_{\rm eff}<25 L_\odot/pc^2$ the DMsd is, in the errors limits, and in the range $1.35 \preceq \Sigma_{\rm eff} \preceq 1.5$, in agreement with the D09 result, after it has smaller values than D09. Concerning MOND's predictions, in the case $\Sigma_{\rm eff}<100 L_\odot/pc^2$, and $1.6 \preceq \log{\Sigma_{\rm eff}} \preceq 2$, MOND is in agreement with D09 prediction, while for smaller values of $\Sigma_{\rm eff}$ it shows a decline towards smaller values of the DMsd, in agreement with the data. 
In the case $\Sigma_{\rm eff}<25 L_\odot/pc^2$, for all values of $\Sigma_{\rm eff}$, DMsd is smaller than D09 prediction, and 
shows a decline towards smaller values of the DMsd, in agreement with data. 

In summary, the SPARC's HSBs tell us that both D09, and MOND correctly predict a constant DMsd, even if the value predicted is smaller than that obtained from data. 
SPARC's LSBs shows a decrease of the DMsd with luminosity and surface brightness (or baryon surface density), in agreement with MOND but not with D09. 

\newpage
{\bf Appendix: DMsd in MOND}\label{sec2}

In this section, we summarize the calculation developed in \citep{DelPopolo2023} to obtain the DMsd in MOND.
Using the MOND formulation by \citep{Bekenstein1984}, the potential $\phi$ is given by a generalization of Poisson equation
\begin{equation}
\nabla [ \mu(|\nabla \phi|/a_0) \nabla \phi ]=4 \pi G \rho
\label{field1}
\end{equation}
where the baryon density is indicated by $\rho$, $\mu(x)$ is the interpolating function, and $a_0$ is the MOND constant. The difference between the MOND acceleration field $\nabla \phi$ and the Newtonian one, interpreted from a Newtonian point of view is explained by the presence of dark matter or as indicated by Milgrom, by "phantom matter" having a density \citep{Milgrom1986}
\begin{equation}
\rho_{\rm p}= \frac{1}{4 \pi G} \Delta \phi- \rho
\end{equation}

Using Eq. (\ref{field1}), we can write
\begin{equation}
\rho_P ({\bf r \rm})=-\frac{1}{4 \pi G a_0} \frac{\mu'}{\mu}  \nabla |\nabla \phi|  \nabla  \phi+\rho ({\bf r}) (1/\mu-1).
\end{equation}

The previous equation can be written as
\begin{equation}
\rho_P=\rho_{P_1}+\rho_{P_2}=\frac{-a_0}{4 \pi G} {\bf e} \cdot \nabla \mathcal{V} ( |\nabla  \phi|/a_0)+(1/\mu-1) \rho.
\label{eq:compl}
\end{equation}
After defining $\mathcal{V}(x)= \int L(x) dx$, $x={\rm g}/a_0$, $L=\frac{\mu'}{\mu} x$, and defining a vector $\bf e$ in the direction of $ \nabla  \phi$, we also have

\begin{equation}
\rho_P=\rho_{P_1}+\rho_{P_2}=\frac{-a_0}{4 \pi G} {\bf e} \cdot \nabla \mathcal{V} ( |\nabla  \phi|/a_0)+(1/\mu-1) \rho,
\label{eq:compl}
\end{equation}
with
\begin{equation}
\rho_{P_1}=\frac{-a_0}{4 \pi G} {\bf e} \nabla  \mathcal{V} (|\nabla  \phi|/a_0),
\label{eq:p1}
\end{equation}
\begin{equation}
\rho_{P_2}=(1/\mu-1) \rho.
\label{eq:p2}
\end{equation}

The integral of Equation~(\ref{eq:compl}) is calculated considering two cases: $x \geq 1$ (acceleration above the MOND universal constant), and $x \leq 1$. 
%
%
Let us consider the second term  
(Equation~\ref{eq:p2}),  $\rho_{P_2}=(1/\mu-1) \rho$. 
We write $a_N=\frac{g_N}{a_0}=({\rm g}/a_0)\mu({\rm g}/a_0)$, where $g_N={\rm g} \mu({\rm g/a_0}) $ is the Newtonian acceleration. 

In the case $n=1$, and $\mu=\frac{x}{(1+x^n)^{1/n}}=\frac{{\rm g}/a_0}{1+{\rm g}/a_0}$, 
multiplying for $\mu$ we have
\begin{equation}
\mu=\frac{\mu {\rm g}/a_0}{\mu {\rm g}/a_0+\mu}=\frac{a_N}{a_N+\mu},
\end{equation} 
solving with respect to $\mu$ we have
\begin{equation}
\mu=-1/2a_N +1/2 \sqrt{a_N^2+4a_N}.
\end{equation}

For a double exponential disk, with~cylindrical radius $R$ and altitude $z$,
we have
\begin{equation}
a{_N} \simeq\frac{GM}{a_0 R{^2}}= \frac{G2 \pi \rho{_0} h{_z} h{_r}{^2}}{a{_0} R{^2}} \simeq \frac{\Sigma{_b}}{\Sigma{_M}} \frac{h{_r}{^2}}{R{^2}}
\end{equation}
being $\Sigma_b$, the baryonic surface density, approximated by an integrated constant volume density ($\Sigma_b\simeq \int \rho_0 dz=\rho_0 h_z$).

The MOND interpolating function becomes
\begin{equation}
\mu=-1/2\,{\frac {\Sigma_{b}\,{{\it h_r}}^{2}}{\Sigma_{M}\,{R}^{2}}}+1/2\,
\sqrt {{\frac {{\Sigma_{b}}^{2}{{\it h_r}}^{4}}{{\Sigma_{M}}^{2}{R}^{4}
}}+4\,{\frac {\Sigma_{b}\,{{\it h_r}}^{2}}{\Sigma_{M}\,{R}^{2}}}}.
\label{eq:muu}
\end{equation}
Then
\begin{equation}
\frac{1-\mu}{\mu}
 \int_{-\infty}^{+\infty} \rho dz = 
2(1-\mu) x \Sigma_M \frac{R^2}{h_r^2} \exp^{-R/h_r}.
\end{equation}
The integral of the first (Eq. 13) term is given by 
\begin{eqnarray}
\int^{+\infty}_{-\infty} \rho_{P_1} dz &=& 2 \int^{\infty}_0 \rho_{P_1} dz=
\nonumber\\
& &
\Sigma_M [\mathcal{V(\infty)}-\mathcal{V}(0)]=
\nonumber\\
& &
\Sigma_M \int^\infty_0 L(x)dx=a \Sigma_M,
\end{eqnarray}
where $\Sigma_M=\frac{a_0}{2 \pi G}$. 

Integrating the Burkert profile D09 obtained a surface density given by
\begin{equation}
\Sigma_0= 2 \int \rho dr= 2 \int \frac{\rho_0 r_0^3}{(r+r_0) (r^2+r^2_0)}=\frac{\pi}{2} \rho_0 r_0=\frac{\pi}{2} \Sigma_c.
\end{equation}
If we call $\Sigma_c^*$ the MOND analog of $\Sigma_c$ we have 
\begin{equation}
\Sigma_c^*=\frac{2\lambda}{\pi } \Sigma_M,
\end{equation}
with $\Sigma_M=138 \frac{a_0}{1.2 \times 10^{-8} \rm cm s^{-2}} M_\odot/pc^2$.

The full integral can be written as
\begin{equation}
F=F_1+F_2=\frac{2}{\pi}\Sigma_M \int_0^x L(x) dx + \frac{2}{\pi} 2(1-\mu) x \Sigma_M \frac{R^2}{h_r^2} \exp^{-R/h_r}.
\label{eq:tot}
\end{equation}

As shown in \citep{DelPopolo2023}, in the case $x \leq 1$ we have 
\begin{eqnarray}
F&=& \frac{2}{\pi} \Sigma_M \arctan{(\sqrt{\frac{\Sigma_b}{\Sigma_M}}\frac{h_r}{R})}+ 
\nonumber\\
& &
\frac{2}{\pi} 2
\frac {R\Sigma_{M}}{{\it h_r}} \left( 1+1/2\,{\frac {\Sigma_{b}\,{{
\it h_r}}^{2}}{\Sigma_{M}\,{R}^{2}}}-1/2\,\sqrt {{\frac {{\Sigma_{b}}^{
2}{{\it h_r}}^{4}}{{\Sigma_{M}}^{2}{R}^{4}}}+4\,{\frac {\Sigma_{b}\,{{
\it h_r}}^{2}}{\Sigma_{M}\,{R}^{2}}}} \right) 
\nonumber\\
& &\times
\sqrt {{\frac {\Sigma_{b}}
{\Sigma_{M}}}}
{{\rm e}^{-{\frac {R}{{\it h_r}}}}}.
\end{eqnarray}

At small $x$ the first term in the right hand side
tend to zero, and $F$ is dominated by the second term. At first order the trend of $F$ is $\sqrt{\Sigma_b \Sigma_M}$.

In the case $x \geq 1$, we have 
\begin{eqnarray}
F &=& \frac{2}{\pi}\Sigma_M \arctan{(\frac{\Sigma_b}{\Sigma_M}\frac{h_r^2}{R^2})}+
\nonumber \\
& &
\frac{4}{\pi}\, \left( 1+1/2\,{\frac {\Sigma_{b}\,{{\it h_r}}^{2}}{\Sigma_{M}\,{R}^
{2}}}-1/2\,\sqrt {{\frac {{\Sigma_{b}}^{2}{{\it h_r}}^{4}}{{\Sigma_{M}}
^{2}{R}^{4}}}+4\,
{\frac {\Sigma_{b}\,{{\it h_r}}^{2}}{\Sigma_{M}\,{R}^{
2}}}} \right)
\nonumber\\
& &\times
\Sigma_{M}\,
{{\rm e}^{-{\frac {R}{{\it h_r}}}}} R^2/h_r^2.
\label{eq:finall}
\end{eqnarray}

For large $x$, the second term 
tends to 0, O(1), and the first one
is $\simeq \Sigma_M$.

%
%

As it comes from the previous results and the plots in \citep{DelPopolo2023} 
there is a double trend of the surface density. For~$R=h_r$, and at~small $\Sigma_b/\Sigma_M$, , the~surface density increases as $\sqrt{\frac{\Sigma_b}{\Sigma_M}} \Sigma_M$. For 
larger $\Sigma_b/\Sigma_M$, the behavior of the surface density 
flattens till when, at~large $\Sigma_b/\Sigma_M$ tends to $\Sigma_M$.   

~\\

{\bf Acknowledgements} 

The author thanks Valerio Pirronello for a careful reading of the manuscript. Financial contribution from the Programma ricerca di Ateneo UNICT 2020-22 linea 2 is graciously acknowl- edged. All authors read and approved the final manuscript.

%

\begin{thebibliography}{}
\expandafter\ifx\csname url\endcsname\relax
  \def\url#1{\texttt{#1}}\fi
\expandafter\ifx\csname urlprefix\endcsname\relax\def\urlprefix{URL }\fi
\expandafter\ifx\csname href\endcsname\relax
  \def\href#1#2{#2} \def\path#1{#1}\fi

\end{thebibliography}


\begin{thebibliography}{10}
\expandafter\ifx\csname url\endcsname\relax
  \def\url#1{\texttt{#1}}\fi
\expandafter\ifx\csname urlprefix\endcsname\relax\def\urlprefix{URL }\fi
\expandafter\ifx\csname href\endcsname\relax
  \def\href#1#2{#2} \def\path#1{#1}\fi

\bibitem{Donato}
F.~{Donato}, G.~{Gentile}, P.~{Salucci}, C.~{Frigerio Martins}, M.~I.
  {Wilkinson}, G.~{Gilmore}, E.~K. {Grebel}, A.~{Koch}, R.~{Wyse}, {A constant
  dark matter halo surface density in galaxies}, \mnras 397 (2009) 1169--1176.
\newblock \href {http://arxiv.org/abs/0904.4054} {\path{arXiv:0904.4054}},
  \href {https://doi.org/10.1111/j.1365-2966.2009.15004.x}
  {\path{doi:10.1111/j.1365-2966.2009.15004.x}}.

\bibitem{Milgrom2009}
M.~{Milgrom}, {The central surface density of `dark haloes' predicted by MOND},
  \mnras 398~(2) (2009) 1023--1026.
\newblock \href {http://arxiv.org/abs/0909.5184} {\path{arXiv:0909.5184}},
  \href {https://doi.org/10.1111/j.1365-2966.2009.15255.x}
  {\path{doi:10.1111/j.1365-2966.2009.15255.x}}.

\bibitem{Spergel}
D.~N. {Spergel}, L.~{Verde}, H.~V. {Peiris}, E.~{Komatsu}, M.~R. {Nolta}, C.~L.
  {Bennett}, M.~{Halpern}, G.~{Hinshaw}, N.~{Jarosik}, A.~{Kogut}, M.~{Limon},
  S.~S. {Meyer}, L.~{Page}, G.~S. {Tucker}, J.~L. {Weiland}, E.~{Wollack},
  E.~L. {Wright}, {First-Year Wilkinson Microwave Anisotropy Probe (WMAP)
  Observations: Determination of Cosmological Parameters}, \apjs 148 (2003)
  175--194.
\newblock \href {http://arxiv.org/abs/astro-ph/0302209}
  {\path{arXiv:astro-ph/0302209}}, \href {https://doi.org/10.1086/377226}
  {\path{doi:10.1086/377226}}.

\bibitem{Kowalski}
M.~{Kowalski}, D.~{Rubin}, G.~{Aldering}, R.~J. {Agostinho}, A.~{Amadon},
  R.~{Amanullah}, C.~{Balland}, K.~{Barbary}, G.~{Blanc}, P.~J. {Challis},
  A.~{Conley}, e.~a. {Connolly}, N.~V., {Improved Cosmological Constraints from
  New, Old, and Combined Supernova Data Sets}, \apj 686 (2008) 749--778.
\newblock \href {http://arxiv.org/abs/0804.4142} {\path{arXiv:0804.4142}},
  \href {https://doi.org/10.1086/589937} {\path{doi:10.1086/589937}}.

\bibitem{Percival}
W.~J. {Percival}, B.~A. {Reid}, D.~J. {Eisenstein}, N.~A. {Bahcall},
  T.~{Budavari}, J.~A. {Frieman}, M.~{Fukugita}, J.~E. {Gunn}, e.~a.
  {Ivezi{\'c}}, {\v Z}., {Baryon acoustic oscillations in the Sloan Digital Sky
  Survey Data Release 7 galaxy sample}, \mnras 401 (2010) 2148--2168.
\newblock \href {http://arxiv.org/abs/0907.1660} {\path{arXiv:0907.1660}},
  \href {https://doi.org/10.1111/j.1365-2966.2009.15812.x}
  {\path{doi:10.1111/j.1365-2966.2009.15812.x}}.

\bibitem{2011ApJS..192...18K}
E.~{Komatsu}, K.~M. {Smith}, J.~{Dunkley}, C.~L. {Bennett}, B.~{Gold},
  G.~{Hinshaw}, N.~{Jarosik}, D.~{Larson}, M.~R. {Nolta}, L.~{Page},
  {Seven-year Wilkinson Microwave Anisotropy Probe (WMAP) Observations:
  Cosmological Interpretation}, \apjs 192~(2) (2011) 18.
\newblock \href {http://arxiv.org/abs/1001.4538} {\path{arXiv:1001.4538}},
  \href {https://doi.org/10.1088/0067-0049/192/2/18}
  {\path{doi:10.1088/0067-0049/192/2/18}}.

\bibitem{DelPopolo2007}
A.~{Del Popolo}, {Dark matter, density perturbations, and structure formation},
  Astronomy Reports 51~(3) (2007) 169--196.
\newblock \href {http://arxiv.org/abs/0801.1091} {\path{arXiv:0801.1091}},
  \href {https://doi.org/10.1134/S1063772907030018}
  {\path{doi:10.1134/S1063772907030018}}.

\bibitem{2013AIPC15482D}
A.~{Del Popolo}, {Non-baryonic dark matter in cosmology} 1548 (2013) 2--63.
\newblock \href {https://doi.org/10.1063/1.4817029}
  {\path{doi:10.1063/1.4817029}}.

\bibitem{2014IJMPD2330005D}
A.~{Del Popolo}, {Nonbaryonic Dark Matter in Cosmology}, International Journal
  of Modern Physics D 23 (2014) 30005.
\newblock \href {http://arxiv.org/abs/1305.0456} {\path{arXiv:1305.0456}},
  \href {https://doi.org/10.1142/S0218271814300055}
  {\path{doi:10.1142/S0218271814300055}}.

\bibitem{Weinberg}
S.~{Weinberg}, {The cosmological constant problem}, Reviews of Modern Physics
  61 (1989) 1--23.
\newblock \href {https://doi.org/10.1103/RevModPhys.61.1}
  {\path{doi:10.1103/RevModPhys.61.1}}.

\bibitem{Astashenok}
A.~V. {Astashenok}, A.~{del Popolo}, {Cosmological measure with volume
  averaging and the vacuum energy problem}, Classical and Quantum Gravity
  29~(8) (2012) 085014.
\newblock \href {http://arxiv.org/abs/1203.2290} {\path{arXiv:1203.2290}},
  \href {https://doi.org/10.1088/0264-9381/29/8/085014}
  {\path{doi:10.1088/0264-9381/29/8/085014}}.

\bibitem{Velten2014}
H.~E.~S. {Velten}, R.~F. {vom Marttens}, W.~{Zimdahl}, {Aspects of the
  cosmological ``coincidence problem''}, European Physical Journal C 74 (2014)
  3160.
\newblock \href {http://arxiv.org/abs/1410.2509} {\path{arXiv:1410.2509}},
  \href {https://doi.org/10.1140/epjc/s10052-014-3160-4}
  {\path{doi:10.1140/epjc/s10052-014-3160-4}}.

\bibitem{DiValentino2021}
E.~{Di Valentino}, O.~{Mena}, S.~{Pan}, L.~{Visinelli}, W.~{Yang},
  A.~{Melchiorri}, D.~F. {Mota}, A.~G. {Riess}, J.~{Silk}, {In the realm of the
  Hubble tension-a review of solutions}, Classical and Quantum Gravity 38~(15)
  (2021) 153001.
\newblock \href {http://arxiv.org/abs/2103.01183} {\path{arXiv:2103.01183}},
  \href {https://doi.org/10.1088/1361-6382/ac086d}
  {\path{doi:10.1088/1361-6382/ac086d}}.

\bibitem{maca}
E.~{Macaulay}, I.~K. {Wehus}, H.~K. {Eriksen}, {Lower Growth Rate from Recent
  Redshift Space Distortion Measurements than Expected from Planck}, Physical
  Review Letters 111~(16) (2013) 161301.
\newblock \href {http://arxiv.org/abs/1303.6583} {\path{arXiv:1303.6583}},
  \href {https://doi.org/10.1103/PhysRevLett.111.161301}
  {\path{doi:10.1103/PhysRevLett.111.161301}}.

\bibitem{raveri}
M.~{Raveri}, {Is there concordance within the concordance $\Lambda$CDM model?},
  ArXiv e-prints (Oct. 2015).
\newblock \href {http://arxiv.org/abs/1510.00688} {\path{arXiv:1510.00688}}.

\bibitem{eriksen}
H.~K. {Eriksen}, F.~K. {Hansen}, A.~J. {Banday}, K.~M. {G{\'o}rski}, P.~B.
  {Lilje}, {Asymmetries in the Cosmic Microwave Background Anisotropy Field},
  \apj 605 (2004) 14--20.
\newblock \href {http://arxiv.org/abs/astro-ph/0307507}
  {\path{arXiv:astro-ph/0307507}}, \href {https://doi.org/10.1086/382267}
  {\path{doi:10.1086/382267}}.

\bibitem{hansen}
F.~K. {Hansen}, A.~J. {Banday}, K.~M. {G{\'o}rski}, {Testing the cosmological
  principle of isotropy: local power-spectrum estimates of the WMAP data},
  \mnras 354 (2004) 641--665.
\newblock \href {http://arxiv.org/abs/astro-ph/0404206}
  {\path{arXiv:astro-ph/0404206}}, \href
  {https://doi.org/10.1111/j.1365-2966.2004.08229.x}
  {\path{doi:10.1111/j.1365-2966.2004.08229.x}}.

\bibitem{jaf}
T.~R. {Jaffe}, A.~J. {Banday}, H.~K. {Eriksen}, K.~M. {G{\'o}rski}, F.~K.
  {Hansen}, {Evidence of Vorticity and Shear at Large Angular Scales in the
  WMAP Data: A Violation of Cosmological Isotropy?}, \apjl 629 (2005) L1--L4.
\newblock \href {http://arxiv.org/abs/astro-ph/0503213}
  {\path{arXiv:astro-ph/0503213}}, \href {https://doi.org/10.1086/444454}
  {\path{doi:10.1086/444454}}.

\bibitem{hot}
J.~{Hoftuft}, H.~K. {Eriksen}, A.~J. {Banday}, K.~M. {G{\'o}rski}, F.~K.
  {Hansen}, P.~B. {Lilje}, {Increasing Evidence for Hemispherical Power
  Asymmetry in the Five-Year WMAP Data}, \apj 699 (2009) 985--989.
\newblock \href {http://arxiv.org/abs/0903.1229} {\path{arXiv:0903.1229}},
  \href {https://doi.org/10.1088/0004-637X/699/2/985}
  {\path{doi:10.1088/0004-637X/699/2/985}}.

\bibitem{planck1}
{Planck Collaboration}, P.~A.~R. {Ade}, N.~{Aghanim}, C.~{Armitage-Caplan},
  M.~{Arnaud}, M.~{Ashdown}, F.~{Atrio-Barandela}, J.~{Aumont},
  C.~{Baccigalupi}, A.~J. {Banday}, et~al., {Planck 2013 results. XXIII.
  Isotropy and statistics of the CMB}, \aap 571 (2014) A23.
\newblock \href {http://arxiv.org/abs/1303.5083} {\path{arXiv:1303.5083}},
  \href {https://doi.org/10.1051/0004-6361/201321534}
  {\path{doi:10.1051/0004-6361/201321534}}.

\bibitem{akrami}
Y.~{Akrami}, Y.~{Fantaye}, A.~{Shafieloo}, H.~K. {Eriksen}, F.~K. {Hansen},
  A.~J. {Banday}, K.~M. {G{\'o}rski}, {Power Asymmetry in WMAP and Planck
  Temperature Sky Maps as Measured by a Local Variance Estimator}, \apjl 784
  (2014) L42.
\newblock \href {http://arxiv.org/abs/1402.0870} {\path{arXiv:1402.0870}},
  \href {https://doi.org/10.1088/2041-8205/784/2/L42}
  {\path{doi:10.1088/2041-8205/784/2/L42}}.

\bibitem{schwa}
D.~J. {Schwarz}, G.~D. {Starkman}, D.~{Huterer}, C.~J. {Copi}, {Is the
  Low-{$l$} Microwave Background Cosmic?}, Physical Review Letters 93~(22)
  (2004) 221301.
\newblock \href {http://arxiv.org/abs/astro-ph/0403353}
  {\path{arXiv:astro-ph/0403353}}, \href
  {https://doi.org/10.1103/PhysRevLett.93.221301}
  {\path{doi:10.1103/PhysRevLett.93.221301}}.

\bibitem{copi}
C.~J. {Copi}, D.~{Huterer}, D.~J. {Schwarz}, G.~D. {Starkman}, {On the
  large-angle anomalies of the microwave sky}, \mnras 367 (2006) 79--102.
\newblock \href {http://arxiv.org/abs/astro-ph/0508047}
  {\path{arXiv:astro-ph/0508047}}, \href
  {https://doi.org/10.1111/j.1365-2966.2005.09980.x}
  {\path{doi:10.1111/j.1365-2966.2005.09980.x}}.

\bibitem{copi1}
C.~J. {Copi}, D.~{Huterer}, D.~J. {Schwarz}, G.~D. {Starkman}, {Uncorrelated
  universe: Statistical anisotropy and the vanishing angular correlation
  function in WMAP years 1 3}, \prd 75~(2) (2007) 023507.
\newblock \href {http://arxiv.org/abs/astro-ph/0605135}
  {\path{arXiv:astro-ph/0605135}}, \href
  {https://doi.org/10.1103/PhysRevD.75.023507}
  {\path{doi:10.1103/PhysRevD.75.023507}}.

\bibitem{copi2}
C.~J. {Copi}, D.~{Huterer}, D.~J. {Schwarz}, G.~D. {Starkman}, {Large-Angle
  Anomalies in the CMB}, Advances in Astronomy 2010 (2010) 847541.
\newblock \href {http://arxiv.org/abs/1004.5602} {\path{arXiv:1004.5602}},
  \href {https://doi.org/10.1155/2010/847541} {\path{doi:10.1155/2010/847541}}.

\bibitem{copi3}
C.~J. {Copi}, D.~{Huterer}, D.~J. {Schwarz}, G.~D. {Starkman}, {Large-scale
  alignments from WMAP and Planck}, \mnras 449 (2015) 3458--3470.
\newblock \href {http://arxiv.org/abs/1311.4562} {\path{arXiv:1311.4562}},
  \href {https://doi.org/10.1093/mnras/stv501}
  {\path{doi:10.1093/mnras/stv501}}.

\bibitem{cruz}
M.~{Cruz}, E.~{Mart{\'{\i}}nez-Gonz{\'a}lez}, P.~{Vielva}, L.~{Cay{\'o}n},
  {Detection of a non-Gaussian spot in WMAP}, \mnras 356 (2005) 29--40.
\newblock \href {http://arxiv.org/abs/astro-ph/0405341}
  {\path{arXiv:astro-ph/0405341}}, \href
  {https://doi.org/10.1111/j.1365-2966.2004.08419.x}
  {\path{doi:10.1111/j.1365-2966.2004.08419.x}}.

\bibitem{cruz1}
M.~{Cruz}, M.~{Tucci}, E.~{Mart{\'{\i}}nez-Gonz{\'a}lez}, P.~{Vielva}, {The
  non-Gaussian cold spot in WilkinsonMicrowaveAnisotropyProbe: significance,
  morphology and foreground contribution}, \mnras 369 (2006) 57--67.
\newblock \href {http://arxiv.org/abs/astro-ph/0601427}
  {\path{arXiv:astro-ph/0601427}}, \href
  {https://doi.org/10.1111/j.1365-2966.2006.10312.x}
  {\path{doi:10.1111/j.1365-2966.2006.10312.x}}.

\bibitem{cruz2}
M.~{Cruz}, L.~{Cay{\'o}n}, E.~{Mart{\'{\i}}nez-Gonz{\'a}lez}, P.~{Vielva},
  J.~{Jin}, {The Non-Gaussian Cold Spot in the 3 Year Wilkinson Microwave
  Anisotropy Probe Data}, \apj 655 (2007) 11--20.
\newblock \href {http://arxiv.org/abs/astro-ph/0603859}
  {\path{arXiv:astro-ph/0603859}}, \href {https://doi.org/10.1086/509703}
  {\path{doi:10.1086/509703}}.

\bibitem{nfw1996}
J.~F. {Navarro}, C.~S. {Frenk}, S.~D.~M. {White}, {The Structure of Cold Dark
  Matter Halos}, \apj 462 (1996) 563.
\newblock \href {http://arxiv.org/abs/astro-ph/9508025}
  {\path{arXiv:astro-ph/9508025}}, \href {https://doi.org/10.1086/177173}
  {\path{doi:10.1086/177173}}.

\bibitem{nfw1997}
J.~F. {Navarro}, C.~S. {Frenk}, S.~D.~M. {White}, {A Universal Density Profile
  from Hierarchical Clustering}, \apj 490 (1997) 493.
\newblock \href {http://arxiv.org/abs/astro-ph/9611107}
  {\path{arXiv:astro-ph/9611107}}, \href {https://doi.org/10.1086/304888}
  {\path{doi:10.1086/304888}}.

\bibitem{Navarro2010}
J.~F. {Navarro}, A.~{Ludlow}, V.~{Springel}, J.~{Wang}, M.~{Vogelsberger},
  S.~D.~M. {White}, A.~{Jenkins}, C.~S. {Frenk}, A.~{Helmi}, {The diversity and
  similarity of simulated cold dark matter haloes}, \mnras 402 (2010) 21--34.
\newblock \href {http://arxiv.org/abs/0810.1522} {\path{arXiv:0810.1522}},
  \href {https://doi.org/10.1111/j.1365-2966.2009.15878.x}
  {\path{doi:10.1111/j.1365-2966.2009.15878.x}}.

\bibitem{Moore1994}
B.~{Moore}, {Evidence against dissipation-less dark matter from observations of
  galaxy haloes}, \nat 370 (1994) 629--631.
\newblock \href {https://doi.org/10.1038/370629a0}
  {\path{doi:10.1038/370629a0}}.

\bibitem{Flores1994}
R.~A. {Flores}, J.~R. {Primack}, {Observational and theoretical constraints on
  singular dark matter halos}, \apjl 427 (1994) L1--L4.
\newblock \href {http://arxiv.org/abs/astro-ph/9402004}
  {\path{arXiv:astro-ph/9402004}}, \href {https://doi.org/10.1086/187350}
  {\path{doi:10.1086/187350}}.

\bibitem{Burkert1995}
A.~{Burkert}, {The Structure of Dark Matter Halos in Dwarf Galaxies}, \apjl 447
  (1995) L25.
\newblock \href {http://arxiv.org/abs/astro-ph/9504041}
  {\path{arXiv:astro-ph/9504041}}, \href {https://doi.org/10.1086/309560}
  {\path{doi:10.1086/309560}}.

\bibitem{deBloketal2003}
W.~J.~G. {de Blok}, A.~{Bosma}, S.~{McGaugh}, {Simulating observations of dark
  matter dominated galaxies: towards the optimal halo profile}, \mnras 340
  (2003) 657--678.
\newblock \href {http://arxiv.org/abs/astro-ph/0212102}
  {\path{arXiv:astro-ph/0212102}}, \href
  {https://doi.org/10.1046/j.1365-8711.2003.06330.x}
  {\path{doi:10.1046/j.1365-8711.2003.06330.x}}.

\bibitem{Swatersetal2003}
R.~A. {Swaters}, B.~F. {Madore}, F.~C. {van den Bosch}, M.~{Balcells}, {The
  Central Mass Distribution in Dwarf and Low Surface Brightness Galaxies}, \apj
  583 (2003) 732--751.
\newblock \href {http://arxiv.org/abs/astro-ph/0210152}
  {\path{arXiv:astro-ph/0210152}}, \href {https://doi.org/10.1086/345426}
  {\path{doi:10.1086/345426}}.

\bibitem{DelPopolo2009}
A.~{Del Popolo}, {The Cusp/Core Problem and the Secondary Infall Model}, \apj
  698 (2009) 2093--2113.
\newblock \href {http://arxiv.org/abs/0906.4447} {\path{arXiv:0906.4447}},
  \href {https://doi.org/10.1088/0004-637X/698/2/2093}
  {\path{doi:10.1088/0004-637X/698/2/2093}}.

\bibitem{DelPopoloKroupa2009}
A.~{Del Popolo}, P.~{Kroupa}, {Density profiles of dark matter haloes on
  galactic and cluster scales}, \aap 502 (2009) 733--747.
\newblock \href {http://arxiv.org/abs/0906.1146} {\path{arXiv:0906.1146}},
  \href {https://doi.org/10.1051/0004-6361/200811404}
  {\path{doi:10.1051/0004-6361/200811404}}.

\bibitem{2012MNRAS419971D}
A.~{Del Popolo}, {Density profile slopes of dwarf galaxies and their
  environment}, \mnras 419 (2012) 971--984.
\newblock \href {http://arxiv.org/abs/1105.0090} {\path{arXiv:1105.0090}},
  \href {https://doi.org/10.1111/j.1365-2966.2011.19754.x}
  {\path{doi:10.1111/j.1365-2966.2011.19754.x}}.

\bibitem{DelPopoloHiotelis2014}
A.~{Del Popolo}, N.~{Hiotelis}, {Cusps and cores in the presence of galactic
  bulges}, \jcap 1 (2014) 47.
\newblock \href {http://arxiv.org/abs/1401.6577} {\path{arXiv:1401.6577}},
  \href {https://doi.org/10.1088/1475-7516/2014/01/047}
  {\path{doi:10.1088/1475-7516/2014/01/047}}.

\bibitem{Klypin1999}
A.~{Klypin}, A.~V. {Kravtsov}, O.~{Valenzuela}, F.~{Prada}, {Where Are the
  Missing Galactic Satellites?}, \apj 522 (1999) 82--92.
\newblock \href {http://arxiv.org/abs/astro-ph/9901240}
  {\path{arXiv:astro-ph/9901240}}, \href {https://doi.org/10.1086/307643}
  {\path{doi:10.1086/307643}}.

\bibitem{Moore1999}
B.~{Moore}, T.~{Quinn}, F.~{Governato}, J.~{Stadel}, G.~{Lake}, {Cold collapse
  and the core catastrophe}, \mnras 310 (1999) 1147--1152.
\newblock \href {http://arxiv.org/abs/astro-ph/9903164}
  {\path{arXiv:astro-ph/9903164}}, \href
  {https://doi.org/10.1046/j.1365-8711.1999.03039.x}
  {\path{doi:10.1046/j.1365-8711.1999.03039.x}}.

\bibitem{GarrisonKimmel2013}
S.~{Garrison-Kimmel}, M.~{Rocha}, M.~{Boylan-Kolchin}, J.~S. {Bullock},
  J.~{Lally}, {Can feedback solve the too-big-to-fail problem?}, \mnras 433~(4)
  (2013) 3539--3546.
\newblock \href {http://arxiv.org/abs/1301.3137} {\path{arXiv:1301.3137}},
  \href {https://doi.org/10.1093/mnras/stt984}
  {\path{doi:10.1093/mnras/stt984}}.

\bibitem{GarrisonKimmel2014}
S.~{Garrison-Kimmel}, M.~{Boylan-Kolchin}, J.~S. {Bullock}, E.~N. {Kirby}, {Too
  big to fail in the Local Group}, \mnras 444~(1) (2014) 222--236.
\newblock \href {http://arxiv.org/abs/1404.5313} {\path{arXiv:1404.5313}},
  \href {https://doi.org/10.1093/mnras/stu1477}
  {\path{doi:10.1093/mnras/stu1477}}.

\bibitem{pawl}
M.~S. {Pawlowski}, B.~{Famaey}, H.~{Jerjen}, D.~{Merritt}, P.~{Kroupa},
  J.~{Dabringhausen}, F.~{L{\"u}ghausen}, D.~A. {Forbes}, G.~{Hensler},
  F.~{Hammer}, M.~{Puech}, S.~{Fouquet}, H.~{Flores}, Y.~{Yang}, {Co-orbiting
  satellite galaxy structures are still in conflict with the distribution of
  primordial dwarf galaxies}, \mnras 442 (2014) 2362--2380.
\newblock \href {http://arxiv.org/abs/1406.1799} {\path{arXiv:1406.1799}},
  \href {https://doi.org/10.1093/mnras/stu1005}
  {\path{doi:10.1093/mnras/stu1005}}.

\bibitem{Sawala2022}
T.~{Sawala}, M.~{Cautun}, C.~{Frenk}, J.~{Helly}, J.~{Jasche}, A.~{Jenkins},
  P.~H. {Johansson}, G.~{Lavaux}, S.~{McAlpine}, M.~{Schaller}, {The Milky
  Way's plane of satellites is consistent with {\ensuremath{\Lambda}}CDM},
  Nature Astronomy (Dec. 2022).
\newblock \href {http://arxiv.org/abs/2205.02860} {\path{arXiv:2205.02860}},
  \href {https://doi.org/10.1038/s41550-022-01856-z}
  {\path{doi:10.1038/s41550-022-01856-z}}.

\bibitem{2003ApJ59849Z}
A.~R. {Zentner}, J.~S. {Bullock}, {Halo Substructure and the Power Spectrum},
  \apj 598 (2003) 49--72.
\newblock \href {http://arxiv.org/abs/astro-ph/0304292}
  {\path{arXiv:astro-ph/0304292}}, \href {https://doi.org/10.1086/378797}
  {\path{doi:10.1086/378797}}.

\bibitem{2000ApJ542622C}
P.~{Col{\'{\i}}n}, V.~{Avila-Reese}, O.~{Valenzuela}, {Substructure and Halo
  Density Profiles in a Warm Dark Matter Cosmology}, \apj 542 (2000) 622--630.
\newblock \href {http://arxiv.org/abs/astro-ph/0004115}
  {\path{arXiv:astro-ph/0004115}}, \href {https://doi.org/10.1086/317057}
  {\path{doi:10.1086/317057}}.

\bibitem{2001ApJ551608S}
J.~{Sommer-Larsen}, A.~{Dolgov}, {Formation of Disk Galaxies: Warm Dark Matter
  and the Angular Momentum Problem}, \apj 551 (2001) 608--623.
\newblock \href {http://arxiv.org/abs/astro-ph/9912166}
  {\path{arXiv:astro-ph/9912166}}, \href {https://doi.org/10.1086/320211}
  {\path{doi:10.1086/320211}}.

\bibitem{2000NewA5103G}
J.~{Goodman}, {Repulsive dark matter}, \na 5 (2000) 103--107.
\newblock \href {http://arxiv.org/abs/astro-ph/0003018}
  {\path{arXiv:astro-ph/0003018}}, \href
  {https://doi.org/10.1016/S1384-1076(00)00015-4}
  {\path{doi:10.1016/S1384-1076(00)00015-4}}.

\bibitem{2000ApJ534L127P}
P.~J.~E. {Peebles}, {Fluid Dark Matter}, \apjl 534 (2000) L127--L129.
\newblock \href {http://arxiv.org/abs/astro-ph/0002495}
  {\path{arXiv:astro-ph/0002495}}, \href {https://doi.org/10.1086/312677}
  {\path{doi:10.1086/312677}}.

\bibitem{1970MNRAS1501B}
H.~A. {Buchdahl}, {Non-linear Lagrangians and cosmological theory}, \mnras 150
  (1970) 1.

\bibitem{1980PhLB9199S}
A.~A. {Starobinsky}, {A new type of isotropic cosmological models without
  singularity}, Physics Letters B 91 (1980) 99--102.
\newblock \href {https://doi.org/10.1016/0370-2693(80)90670-X}
  {\path{doi:10.1016/0370-2693(80)90670-X}}.

\bibitem{1983ApJ270365M}
M.~{Milgrom}, {A modification of the Newtonian dynamics as a possible
  alternative to the hidden mass hypothesis}, \apj 270 (1983) 365--370.
\newblock \href {https://doi.org/10.1086/161130} {\path{doi:10.1086/161130}}.

\bibitem{1983ApJ270371M}
M.~{Milgrom}, {A modification of the Newtonian dynamics - Implications for
  galaxies}, \apj 270 (1983) 371--389.
\newblock \href {https://doi.org/10.1086/161131} {\path{doi:10.1086/161131}}.

\bibitem{Ferraro2012}
R.~{Ferraro}, {f(R) and f(T) theories of modified gravity} 1471 (2012)
  103--110.
\newblock \href {http://arxiv.org/abs/1204.6273} {\path{arXiv:1204.6273}},
  \href {https://doi.org/10.1063/1.4756821} {\path{doi:10.1063/1.4756821}}.

\bibitem{1996MNRAS283L72N}
J.~F. {Navarro}, V.~R. {Eke}, C.~S. {Frenk}, {The cores of dwarf galaxy
  haloes}, \mnras 283 (1996) L72--L78.
\newblock \href {http://arxiv.org/abs/astro-ph/9610187}
  {\path{arXiv:astro-ph/9610187}}.

\bibitem{Gelato1999}
S.~{Gelato}, J.~{Sommer-Larsen}, {On DDO 154 and cold dark matter halo
  profiles}, \mnras 303 (1999) 321--328.
\newblock \href {http://arxiv.org/abs/astro-ph/9806289}
  {\path{arXiv:astro-ph/9806289}}, \href
  {https://doi.org/10.1046/j.1365-8711.1999.02223.x}
  {\path{doi:10.1046/j.1365-8711.1999.02223.x}}.

\bibitem{Read2005}
J.~I. {Read}, G.~{Gilmore}, {Mass loss from dwarf spheroidal galaxies: the
  origins of shallow dark matter cores and exponential surface brightness
  profiles}, \mnras 356 (2005) 107--124.
\newblock \href {http://arxiv.org/abs/astro-ph/0409565}
  {\path{arXiv:astro-ph/0409565}}, \href
  {https://doi.org/10.1111/j.1365-2966.2004.08424.x}
  {\path{doi:10.1111/j.1365-2966.2004.08424.x}}.

\bibitem{Mashchenko2006}
S.~{Mashchenko}, H.~M.~P. {Couchman}, J.~{Wadsley}, {The removal of cusps from
  galaxy centres by stellar feedback in the early Universe}, \nat 442 (2006)
  539--542.
\newblock \href {http://arxiv.org/abs/astro-ph/0605672}
  {\path{arXiv:astro-ph/0605672}}, \href {https://doi.org/10.1038/nature04944}
  {\path{doi:10.1038/nature04944}}.

\bibitem{Governato2010}
F.~{Governato}, C.~{Brook}, L.~{Mayer}, A.~{Brooks}, G.~{Rhee}, J.~{Wadsley},
  P.~{Jonsson}, B.~{Willman}, G.~{Stinson}, T.~{Quinn}, P.~{Madau}, {Bulgeless
  dwarf galaxies and dark matter cores from supernova-driven outflows}, \nat
  463 (2010) 203--206.
\newblock \href {http://arxiv.org/abs/0911.2237} {\path{arXiv:0911.2237}},
  \href {https://doi.org/10.1038/nature08640} {\path{doi:10.1038/nature08640}}.

\bibitem{El-Zant2001}
A.~{El-Zant}, I.~{Shlosman}, Y.~{Hoffman}, {Dark Halos: The Flattening of the
  Density Cusp by Dynamical Friction}, \apj 560 (2001) 636--643.
\newblock \href {http://arxiv.org/abs/astro-ph/0103386}
  {\path{arXiv:astro-ph/0103386}}, \href {https://doi.org/10.1086/322516}
  {\path{doi:10.1086/322516}}.

\bibitem{El-Zant2004}
A.~A. {El-Zant}, Y.~{Hoffman}, J.~{Primack}, F.~{Combes}, I.~{Shlosman},
  {Flat-cored Dark Matter in Cuspy Clusters of Galaxies}, \apjl 607 (2004)
  L75--L78.
\newblock \href {http://arxiv.org/abs/astro-ph/0309412}
  {\path{arXiv:astro-ph/0309412}}, \href {https://doi.org/10.1086/421938}
  {\path{doi:10.1086/421938}}.

\bibitem{2008ApJ685L105R}
E.~{Romano-D{\'{\i}}az}, I.~{Shlosman}, Y.~{Hoffman}, C.~{Heller}, {Erasing
  Dark Matter Cusps in Cosmological Galactic Halos with Baryons}, \apjl 685
  (2008) L105--L108.
\newblock \href {http://arxiv.org/abs/0808.0195} {\path{arXiv:0808.0195}},
  \href {https://doi.org/10.1086/592687} {\path{doi:10.1086/592687}}.

\bibitem{Cole2011}
D.~R. {Cole}, W.~{Dehnen}, M.~I. {Wilkinson}, {Weakening dark matter cusps by
  clumpy baryonic infall}, \mnras 416 (2011) 1118--1134.
\newblock \href {http://arxiv.org/abs/1105.4050} {\path{arXiv:1105.4050}},
  \href {https://doi.org/10.1111/j.1365-2966.2011.19110.x}
  {\path{doi:10.1111/j.1365-2966.2011.19110.x}}.

\bibitem{Saburova2014}
A.~{Saburova}, A.~{Del Popolo}, {On the surface density of dark matter haloes},
  \mnras 445~(4) (2014) 3512--3524.
\newblock \href {http://arxiv.org/abs/1410.3052} {\path{arXiv:1410.3052}},
  \href {https://doi.org/10.1093/mnras/stu1957}
  {\path{doi:10.1093/mnras/stu1957}}.

\bibitem{Kormendy2004}
J.~{Kormendy}, K.~C. {Freeman}, {Scaling Laws for Dark Matter Halos in
  Late-Type and Dwarf Spheroidal Galaxies} 220 (2004) 377.
\newblock \href {http://arxiv.org/abs/astro-ph/0407321}
  {\path{arXiv:astro-ph/0407321}}.

\bibitem{DelPopolo2023}
A.~{Del Popolo}, M.~{Le Delliou}, {Surface Density of Disk Galaxies in MOND},
  Universe 9~(1) (2023) 32.
\newblock \href {https://doi.org/10.3390/universe9010032}
  {\path{doi:10.3390/universe9010032}}.

\bibitem{Gentile2004}
G.~{Gentile}, P.~{Salucci}, U.~{Klein}, D.~{Vergani}, P.~{Kalberla}, {The cored
  distribution of dark matter in spiral galaxies}, \mnras 351 (2004) 903--922.
\newblock \href {http://arxiv.org/abs/astro-ph/0403154}
  {\path{arXiv:astro-ph/0403154}}, \href
  {https://doi.org/10.1111/j.1365-2966.2004.07836.x}
  {\path{doi:10.1111/j.1365-2966.2004.07836.x}}.

\bibitem{Gentile2007}
G.~{Gentile}, P.~{Salucci}, U.~{Klein}, G.~L. {Granato}, {NGC 3741: the dark
  halo profile from the most extended rotation curve}, \mnras 375 (2007)
  199--212.
\newblock \href {http://arxiv.org/abs/astro-ph/0611355}
  {\path{arXiv:astro-ph/0611355}}, \href
  {https://doi.org/10.1111/j.1365-2966.2006.11283.x}
  {\path{doi:10.1111/j.1365-2966.2006.11283.x}}.

\bibitem{Simon2005}
J.~D. {Simon}, A.~D. {Bolatto}, A.~{Leroy}, L.~{Blitz}, E.~L. {Gates},
  {High-Resolution Measurements of the Halos of Four Dark Matter-Dominated
  Galaxies: Deviations from a Universal Density Profile}, \apj 621 (2005)
  757--776.
\newblock \href {http://arxiv.org/abs/astro-ph/0412035}
  {\path{arXiv:astro-ph/0412035}}, \href {https://doi.org/10.1086/427684}
  {\path{doi:10.1086/427684}}.

\bibitem{THINGS}
W.~J.~G. {de Blok}, F.~{Walter}, E.~{Brinks}, C.~{Trachternach}, S.-H. {Oh},
  R.~C. {Kennicutt}, Jr., {High-Resolution Rotation Curves and Galaxy Mass
  Models from THINGS}, \aj 136 (2008) 2648--2719.
\newblock \href {http://arxiv.org/abs/0810.2100} {\path{arXiv:0810.2100}},
  \href {https://doi.org/10.1088/0004-6256/136/6/2648}
  {\path{doi:10.1088/0004-6256/136/6/2648}}.

\bibitem{2012MNRAS42438D}
A.~{Del Popolo}, {On the density-profile slope of clusters of galaxies}, \mnras
  424 (2012) 38--51.
\newblock \href {http://arxiv.org/abs/1204.4439} {\path{arXiv:1204.4439}},
  \href {https://doi.org/10.1111/j.1365-2966.2012.21141.x}
  {\path{doi:10.1111/j.1365-2966.2012.21141.x}}.

\bibitem{Boyarsky}
A.~{Boyarsky}, O.~{Ruchayskiy}, D.~{Iakubovskyi}, A.~V. {Maccio'},
  D.~{Malyshev}, {New evidence for dark matter}, ArXiv e-prints (Nov. 2009).
\newblock \href {http://arxiv.org/abs/0911.1774} {\path{arXiv:0911.1774}}.

\bibitem{CardoneTortora2010}
V.~F. {Cardone}, C.~{Tortora}, {Dark matter scaling relations in intermediate z
  haloes}, \mnras 409 (2010) 1570--1576.
\newblock \href {http://arxiv.org/abs/1007.3673} {\path{arXiv:1007.3673}},
  \href {https://doi.org/10.1111/j.1365-2966.2010.17398.x}
  {\path{doi:10.1111/j.1365-2966.2010.17398.x}}.

\bibitem{Napolitano2010}
N.~R. {Napolitano}, A.~J. {Romanowsky}, C.~{Tortora}, {The central dark matter
  content of early-type galaxies: scaling relations and connections with star
  formation histories}, \mnras 405 (2010) 2351--2371.
\newblock \href {http://arxiv.org/abs/1003.1716} {\path{arXiv:1003.1716}},
  \href {https://doi.org/10.1111/j.1365-2966.2010.16710.x}
  {\path{doi:10.1111/j.1365-2966.2010.16710.x}}.

\bibitem{2013MNRAS4291080D}
A.~{Del Popolo}, V.~F. {Cardone}, G.~{Belvedere}, {Surface density of dark
  matter haloes on galactic and cluster scales}, \mnras 429 (2013) 1080--1087.
\newblock \href {http://arxiv.org/abs/1212.6797} {\path{arXiv:1212.6797}},
  \href {https://doi.org/10.1093/mnras/sts389}
  {\path{doi:10.1093/mnras/sts389}}.

\bibitem{CardoneDelPopolo2012}
V.~F. {Cardone}, A.~{Del Popolo}, {Newtonian acceleration scales in spiral
  galaxies}, \mnras 427 (2012) 3176--3187.
\newblock \href {http://arxiv.org/abs/1209.1524} {\path{arXiv:1209.1524}},
  \href {https://doi.org/10.1111/j.1365-2966.2012.21982.x}
  {\path{doi:10.1111/j.1365-2966.2012.21982.x}}.

\bibitem{Zhou2020}
Y.~{Zhou}, A.~{Del Popolo}, Z.~{Chang}, {On the absence of a universal surface
  density, and a maximum Newtonian acceleration in dark matter haloes:
  Consequences for MOND}, Physics of the Dark Universe 28 (2020) 100468.
\newblock \href {http://arxiv.org/abs/2008.04065} {\path{arXiv:2008.04065}},
  \href {https://doi.org/10.1016/j.dark.2020.100468}
  {\path{doi:10.1016/j.dark.2020.100468}}.

\bibitem{Lelli2016}
F.~Lelli, S.~S. McGaugh, J.~M. Schombert, {SPARC: Mass Models for 175 Disk
  Galaxies with Spitzer Photometry and Accurate Rotation Curves}, Astron. J.
  152 (2016) 157.
\newblock \href {http://arxiv.org/abs/1606.09251} {\path{arXiv:1606.09251}},
  \href {https://doi.org/10.3847/0004-6256/152/6/157}
  {\path{doi:10.3847/0004-6256/152/6/157}}.

\bibitem{Li:2018rnd}
P.~Li, F.~Lelli, S.~S. McGaugh, N.~Starkman, J.~M. Schombert, {A constant
  characteristic volume density of dark matter haloes from SPARC rotation curve
  fits}, Mon. Not. Roy. Astron. Soc. 482~(4) (2019) 5106--5124.
\newblock \href {http://arxiv.org/abs/1811.00553} {\path{arXiv:1811.00553}},
  \href {https://doi.org/10.1093/mnras/sty2968}
  {\path{doi:10.1093/mnras/sty2968}}.

\bibitem{Bekenstein1984}
J.~{Bekenstein}, M.~{Milgrom}, {Does the missing mass problem signal the
  breakdown of Newtonian gravity?}, \apj 286 (1984) 7--14.
\newblock \href {https://doi.org/10.1086/162570} {\path{doi:10.1086/162570}}.

\bibitem{Milgrom1986}
M.~{Milgrom}, {Can the Hidden Mass Be Negative?}, \apj 306 (1986) 9.
\newblock \href {https://doi.org/10.1086/164314} {\path{doi:10.1086/164314}}.

\end{thebibliography}
%

\setlength\LTcapwidth{\textheight}
\scriptsize
	\begin{longtable}{cccc}
		\caption{\label{sample} In the table, the first column gives the name of the galaxy, the second one the best fitting values of the DM surface density (obtained in \citep{Zhou2020}), and the relative errors, the third column the luminosity at 3.6 $\rm \mu m$, and the relative errors given in the SPARC's webpage, and the fourth column the effective surface brightness given in SPARC's webpage.}\\
		\hline	
    Name  & $\log{(\rho_0 r_0)} (M_\odot/pc^3)$ & \multicolumn{1}{l}{L[3.6 $\mu m$] $(10^9 L_{\odot})$ } & \multicolumn{1}{l}{$\Sigma_{eff}(L_{\odot}/pc^2)$} \\
\hline  
\endfirsthead
\caption{Continued}\\
		\hline
		Name  & $\log{(\rho_0 r_0)} (M_\odot/pc^3)$ & \multicolumn{1}{l}{L[3.6 $\mu m$] $(10^9 L_{\odot})$ } & \multicolumn{1}{l}{$\Sigma_{eff}(L_{\odot}/pc^2)$} \\
		\hline
	\endhead
    UGC02487 & 2.33 $\pm$0.23 & 489.955$\pm$4.061 & 818.14 \\
    UGC02885 & 2.24$\pm$0.11 & 403.525$\pm$4.088 & 421.53 \\
    NGC6195 & 2.18$\pm$0.1 & 391.076$\pm$6.123 & 686.8 \\
    UGC11455 & 3.01$\pm$0.11 & 374.322$\pm$3.792 & 571.26 \\
    NGC5371 & 2.13$\pm$0.09 & 340.393$\pm$1.881 & 560.83 \\
    NGC2955 & 2.57$\pm$0.11 & 319.422$\pm$4.413 & 974.61 \\
    NGC0801 & 1.82$\pm$0.06 & 312.57$\pm$3.455 & 818.14 \\
    ESO563-G021 & 3.28$\pm$0.13 & 311.177$\pm$2.579 & 429.37 \\
    UGC09133 & 1.92$\pm$0.06 & 282.926$\pm$2.345 & 1284.78 \\
    UGC02953 & 3.64$\pm$0.06 & 259.518$\pm$0.717 & 1544.65 \\
    NGC7331 & 2.4$\pm$0.06 & 250.631$\pm$0.693 & 2470.76 \\
    NGC3992 & 2.82$\pm$0.18 & 226.932$\pm$0.836 & 360.44 \\
    NGC6674 & 1.69$\pm$0.07 & 214.654$\pm$1.977 & 560.83 \\
    NGC5985 & 3.37$\pm$0.15 & 208.728$\pm$1.538 & 288.95 \\
    NGC2841 & 2.25$\pm$0.04 & 188.121$\pm$0.52 & 983.63 \\
    IC4202 & 3.55$\pm$0.06 & 179.749$\pm$3.311 & 391.59 \\
    NGC5005 & 2.67$\pm$0.21 & 178.72$\pm$0.494 & 1789.91 \\
    NGC5907 & 2.93$\pm$0.04 & 175.425$\pm$0.646 & 449.6 \\
    UGC05253 & 2.77$\pm$0.09 & 171.582$\pm$0.79 & 1488.78 \\
    NGC5055 & 2.22$\pm$0.06 & 152.922$\pm$0.282 & 1383.02 \\
    NGC2998 & 1.94$\pm$0.09 & 150.902$\pm$2.085 & 479.55 \\
    UGC11914 & 2.95$\pm$0.08 & 150.028$\pm$0.553 & 2425.66 \\
    NGC3953 & 2.79$\pm$0.45 & 141.301$\pm$0.521 & 587.26 \\
    UGC12506 & 2.9$\pm$0.07 & 139.571$\pm$3.214 & 144.82 \\
    NGC0891 & 2.82$\pm$0.06 & 138.34$\pm$0.255 & 1617.45 \\
    UGC06614 & 2.22$\pm$0.17 & 124.35$\pm$2.52 & 1461.61 \\
    UGC02916 & 2.68$\pm$0.08 & 124.153$\pm$1.83 & 2425.66 \\
    UGC03205 & 3.06$\pm$0.09 & 113.642$\pm$1.361 & 587.26 \\
    NGC5033 & 2.65$\pm$0.09 & 110.509$\pm$0.407 & 2036.24 \\
    NGC4088 & 2.21$\pm$0.15 & 107.286$\pm$0.494 & 457.96 \\
    NGC4157 & 2.12$\pm$0.1 & 105.62$\pm$0.486 & 848.85 \\
    UGC03546 & 2.57$\pm$0.11 & 101.336$\pm$0.747 & 2403.42 \\
    UGC06787 & 1.9$\pm$0.06 & 98.256$\pm$0.543 & 1891.6 \\
    NGC4051 & 2.44$\pm$0.37 & 95.268$\pm$0.439 & 319.76 \\
    NGC4217 & 2.98$\pm$0.16 & 85.299$\pm$0.393 & 483.98 \\
    NGC3521 & 2.35$\pm$0.1 & 84.836$\pm$0.156 & 2232.7 \\
    NGC2903 & 2.82$\pm$0.2 & 81.863$\pm$0.151 & 632.16 \\
    NGC2683 & 2.67$\pm$0.19 & 80.415$\pm$0.222 & 1139.8 \\
    NGC4013 & 2.03$\pm$0.05 & 79.094$\pm$0.364 & 746.16 \\
    NGC7814 & 2.71$\pm$0.06 & 74.529$\pm$0.343 & 2709.13 \\
    UGC06786 & 2.43$\pm$0.09 & 73.407$\pm$0.676 & 992.73 \\
    NGC3877 & 2.88$\pm$0.11 & 72.535$\pm$0.401 & 598.18 \\
    NGC0289 & 1.84$\pm$0.12 & 72.065$\pm$0.465 & 1587.93 \\
    NGC1090 & 2.82$\pm$0.22 & 72.045$\pm$0.796 & 283.68 \\
    NGC3726 & 1.96$\pm$0.12 & 70.234$\pm$0.388 & 198.08 \\
    UGC09037 & 2.33$\pm$0.1 & 68.614$\pm$1.769 & 334.83 \\
    NGC6946 & 2.31$\pm$0.16 & 66.173$\pm$0.122 & 571.26 \\
    NGC4100 & 2.79$\pm$0.11 & 59.394$\pm$0.328 & 388 \\
    NGC3893 & 2.68$\pm$0.16 & 58.525$\pm$0.377 & 1573.37 \\
    UGC06973 & 2.98$\pm$0.11 & 53.87$\pm$0.347 & 3317.65 \\
    ESO079-G014 & 2.56$\pm$0.15 & 51.733$\pm$0.524 & 157.34 \\
    UGC08699 & 2.17$\pm$0.09 & 50.302$\pm$0.695 & 2191.95 \\
    NGC4138 & 2.82$\pm$0.29 & 44.111$\pm$0.284 & 1909.11 \\
    NGC3198 & 2.15$\pm$0.06 & 38.279$\pm$0.212 & 178.99 \\
    NGC3949 & 2.86$\pm$0.23 & 38.067$\pm$0.28 & 1058.84 \\
    NGC6015 & 1.73$\pm$0.1 & 32.129$\pm$0.237 & 331.76 \\
    NGC3917 & 2.44$\pm$0.1 & 21.966$\pm$0.202 & 112.94 \\
    NGC4085 & 2.55$\pm$0.15 & 21.724$\pm$0.2 & 856.7 \\
    NGC4389 & 3.03$\pm$0.22 & 21.328$\pm$0.216 & 207.41 \\
    NGC4559 & 2.05$\pm$0.15 & 19.377$\pm$0.107 & 211.27 \\
    NGC3769 & 2.25$\pm$0.12 & 18.679$\pm$0.189 & 626.37 \\
    NGC4010 & 2.27$\pm$0.11 & 17.193$\pm$0.19 & 14.75 \\
    NGC3972 & 2.48$\pm$0.12 & 14.353$\pm$0.172 & 164.75 \\
    UGC03580 & 2.33$\pm$0.07 & 13.266$\pm$0.195 & 614.94 \\
    NGC6503 & 2.26$\pm$0.03 & 12.845$\pm$0.059 & 774.16 \\
    UGC11557 & 2.36$\pm$0.53 & 12.101$\pm$0.212 & 106.86 \\
    UGC00128 & 2.18$\pm$0.1 & 12.02$\pm$0.565 & 20.36 \\
    F579-V1 & 2.46$\pm$0.4 & 11.848$\pm$0.742 & 56.6 \\
    NGC4183 & 2.38$\pm$0.18 & 10.838$\pm$0.15 & 86.46 \\
    F571-8 & 2.79$\pm$0.14 & 10.164$\pm$0.412 & 825.71 \\
    NGC2403 & 2.34$\pm$0.04 & 10.041$\pm$0.028 & 341.06 \\
    UGC06930 & 2.29$\pm$0.25 & 8.932$\pm$0.14 & 73.93 \\
    F568-3 & 2.31$\pm$0.29 & 8.346$\pm$0.592 & 23.81 \\
    UGC01230 & 2.5$\pm$0.54 & 7.62$\pm$0.379 & 28.63 \\
    NGC0247 & 2.06$\pm$0.18 & 7.332$\pm$0.027 & 33.79 \\
    NGC7793 & 2.01$\pm$0.16 & 7.05$\pm$0.026 & 233.79 \\
    UGC06917 & 2.33$\pm$0.1 & 6.832$\pm$0.12 & 53.07 \\
    NGC1003 & 1.7$\pm$0.09 & 6.82$\pm$0.075 & 140.87 \\
    F574-1 & 2.31$\pm$0.12 & 6.537$\pm$0.596 & 29.98 \\
    F568-1 & 2.54$\pm$0.21 & 6.252$\pm$0.564 & 20.36 \\
    UGC06983 & 2.39$\pm$0.1 & 5.298$\pm$0.102 & 53.56 \\
    UGC05986 & 2.43$\pm$0.15 & 4.695$\pm$0.048 & 76.71 \\
    NGC0055 & 1.94$\pm$0.04 & 4.628$\pm$0.013 & 54.55 \\
    ESO116-G012 & 2.33$\pm$0.16 & 4.292$\pm$0.071 & 90.54 \\
    UGC07323 & 2.01$\pm$0.25 & 4.109$\pm$0.042 & 61.49 \\
    UGC05005 & 1.84$\pm$0.31 & 4.1$\pm$0.283 & 25.63 \\
    F561-1 & 1.9$\pm$0.59 & 4.077$\pm$0.327 & 22.12 \\
    NGC0024 & 2.8$\pm$0.35 & 3.889$\pm$0.036 & 151.65 \\
    F568-V1 & 2.53$\pm$0.28 & 3.825$\pm$0.384 & 31.39 \\
    UGC06628 & 2.16$\pm$0.6 & 3.739$\pm$0.076 & 34.74 \\
    UGC02455 & 3.63$\pm$0.31 & 3.649$\pm$0.034 & 254 \\
    UGC07089 & 1.85$\pm$0.2 & 3.585$\pm$0.089 & 37.4 \\
    UGC05999 & 2.24$\pm$0.54 & 3.384$\pm$0.231 & 22.95 \\
    NGC2976 & 2.46$\pm$0.26 & 3.371$\pm$0.019 & 308.2 \\
    UGC05750 & 1.8$\pm$0.21 & 3.336$\pm$0.264 & 6.87 \\
    NGC0100 & 2.13$\pm$0.2 & 3.232$\pm$0.063 & 64.99 \\
    UGC00634 & 2.08$\pm$0.28 & 2.989$\pm$0.146 & 26.11 \\
    F563-V2 & 2.62$\pm$0.4 & 2.986$\pm$0.267 & 23.6 \\
    NGC5585 & 2$\pm$0.09 & 2.943$\pm$0.033 & 90.54 \\
    NGC0300 & 2.11$\pm$0.23 & 2.922$\pm$0.008 & 147.51 \\
    UGC06923 & 2.17$\pm$0.18 & 2.89$\pm$0.077 & 166.28 \\
    F574-2 & 0.74$\pm$0.5 & 2.877$\pm$0.384 & 10.89 \\
    UGC07125 & 1.77$\pm$0.17 & 2.712$\pm$0.08 & 28.11 \\
    UGC07524 & 2.04$\pm$0.07 & 2.436$\pm$0.025 & 29.71 \\
    UGC06399 & 2.21$\pm$0.09 & 2.296$\pm$0.072 & 30.82 \\
    UGC07151 & 2.28$\pm$0.09 & 2.284$\pm$0.025 & 76.71 \\
    F567-2 & 1.87$\pm$0.66 & 2.134$\pm$0.305 & 11.5 \\
    UGC04325 & 2.67$\pm$0.17 & 2.026$\pm$0.035 & 41 \\
    UGC00191 & 2.21$\pm$0.39 & 2.004$\pm$0.063 & 50.21 \\
    F563-1 & 2.38$\pm$0.25 & 1.903$\pm$0.17 & 14.22 \\
    F571-V1 & 1.97$\pm$0.42 & 1.849$\pm$0.267 & 15.3 \\
    UGC07261 & 2.44$\pm$0.47 & 1.753$\pm$0.048 & 39.16 \\
    UGC10310 & 2.21$\pm$0.31 & 1.741$\pm$0.053 & 28.37 \\
    UGC02259 & 2.51$\pm$0.17 & 1.725$\pm$0.038 & 47.08 \\
    F583-4 & 2.01$\pm$0.25 & 1.715$\pm$0.185 & 24.94 \\
    UGC12732 & 1.73$\pm$0.2 & 1.667$\pm$0.048 & 26.84 \\
    UGC06818 & 2.15$\pm$0.22 & 1.588$\pm$0.057 & 56.08 \\
    UGC04499 & 2.1$\pm$0.21 & 1.552$\pm$0.043 & 33.79 \\
    F563-V1 & 1$\pm$0.41 & 1.54$\pm$0.165 & 9.75 \\
    UGC06667 & 2.26$\pm$0.07 & 1.397$\pm$0.066 & 18.06 \\
    UGC02023 & 2.73$\pm$0.6 & 1.308$\pm$0.033 & 27.59 \\
    UGC04278 & 2.08$\pm$0.15 & 1.307$\pm$0.026 & 34.11 \\
    UGC12632 & 2.04$\pm$0.18 & 1.301$\pm$0.03 & 13.09 \\
    UGC08286 & 2.46$\pm$0.03 & 1.255$\pm$0.018 & 39.52 \\
    UGC07399 & 2.6$\pm$0.16 & 1.156$\pm$0.024 & 113.98 \\
    NGC4214 & 1.9$\pm$0.22 & 1.141$\pm$0.008 & 363.77 \\
    UGC05414 & 1.98$\pm$0.24 & 1.123$\pm$0.028 & 32.87 \\
    UGC08490 & 2.31$\pm$0.14 & 1.017$\pm$0.012 & 124.98 \\
    IC2574 & 1.69$\pm$0.12 & 1.016$\pm$0.012 & 15.88 \\
    UGC06446 & 2.29$\pm$0.16 & 0.988$\pm$0.032 & 37.05 \\
    F583-1 & 2.13$\pm$0.15 & 0.986$\pm$0.093 & 11.19 \\
    UGC11820 & 1.3$\pm$0.18 & 0.97$\pm$0.047 & 20.18 \\
    UGC07690 & 2.45$\pm$0.35 & 0.858$\pm$0.018 & 184.01 \\
    UGC04305 & 2.31$\pm$0.44 & 0.736$\pm$0.007 & 76.71 \\
    NGC2915 & 2.46$\pm$0.07 & 0.641$\pm$0.008 & 347.4 \\
    UGC05716 & 1.62$\pm$0.12 & 0.588$\pm$0.042 & 27.59 \\
    UGC05829 & 1.77$\pm$0.4 & 0.564$\pm$0.019 & 10.59 \\
    F565-V2 & 2.06$\pm$0.18 & 0.559$\pm$0.098 & 6.93 \\
    DDO161 & 1.65$\pm$0.14 & 0.548$\pm$0.015 & 20.74 \\
    DDO170 & 1.99$\pm$0.2 & 0.543$\pm$0.03 & 9.39 \\
    NGC1705 & 2.71$\pm$0.11 & 0.533$\pm$0.01 & 347.4 \\
    UGC05721 & 2.63$\pm$0.16 & 0.531$\pm$0.011 & 233.79 \\
    UGC08837 & 2.5$\pm$0.24 & 0.501$\pm$0.015 & 15.73 \\
    UGC07603 & 2.32$\pm$0.19 & 0.376$\pm$0.008 & 81.81 \\
    UGC00891 & 1.89$\pm$0.18 & 0.374$\pm$0.017 & 19.09 \\
    UGC01281 & 1.99$\pm$0.04 & 0.353$\pm$0.009 & 13.83 \\
    UGC09992 & 2.03$\pm$0.52 & 0.336$\pm$0.017 & 20.18 \\
    D512-2 & 1.79$\pm$0.3 & 0.325$\pm$0.022 & 9.22 \\
    UGC00731 & 2.06$\pm$0.11 & 0.323$\pm$0.019 & 25.63 \\
    UGC08550 & 2.14$\pm$0.17 & 0.289$\pm$0.009 & 45.38 \\
    UGC07608 & 2.18$\pm$0.47 & 0.264$\pm$0.012 & 16.48 \\
    NGC4068 & 2.57$\pm$0.32 & 0.236$\pm$0.005 & 30.26 \\
    NGC2366 & 1.9$\pm$0.06 & 0.236$\pm$0.005 & 31.98 \\
    UGC05918 & 1.95$\pm$0.21 & 0.233$\pm$0.011 & 5.36 \\
    D631-7 & 1.89$\pm$0.06 & 0.196$\pm$0.009 & 20.93 \\
    NGC3109 & 2$\pm$0.04 & 0.194$\pm$0.002 & 11.4 \\
    UGCA281 & 2.03$\pm$0.2 & 0.194$\pm$0.007 & 12.5 \\
    DDO168 & 2.12$\pm$0.08 & 0.191$\pm$0.005 & 18.23 \\
    DDO064 & 2.01$\pm$0.28 & 0.157$\pm$0.007 & 17.41 \\
    PGC51017 & 1.64$\pm$0.39 & 0.155$\pm$0.014 & 15.03 \\
    UGCA442 & 1.96$\pm$0.08 & 0.14$\pm$0.005 & 7.6 \\
    UGC07866 & 1.77$\pm$0.37 & 0.124$\pm$0.004 & 21.92 \\
    UGC07232 & 3.21$\pm$0.3 & 0.113$\pm$0.002 & 83.34 \\
    UGC07559 & 1.65$\pm$0.37 & 0.109$\pm$0.004 & 17.9 \\
    NGC6789 & 2.87$\pm$0.32 & 0.1$\pm$0.003 & 59.27 \\
    KK98-251 & 1.52$\pm$0.32 & 0.085$\pm$0.007 & 7.89 \\
    UGC05764 & 2.36$\pm$0.15 & 0.085$\pm$0.006 & 9.31 \\
    CamB  & 2.55$\pm$0.28 & 0.075$\pm$0.003 & 7.89 \\
    ESO444-G084 & 2.27$\pm$0.22 & 0.071$\pm$0.003 & 19.81 \\
    DDO154 & 1.86$\pm$0.04 & 0.053$\pm$0.002 & 19.99 \\
    UGC07577 & 2.18$\pm$0.41 & 0.045$\pm$0.002 & 11.94 \\
    D564-8 & 1.46$\pm$0.25 & 0.033$\pm$0.004 & 10.11 \\
    NGC3741 & 1.82$\pm$0.04 & 0.028$\pm$0.001 & 42.94 \\
    UGC04483 & 1.84$\pm$0.2 & 0.013$\pm$0.001 & 29.71 \\
    UGCA444 & 1.83$\pm$0.05 & 0.012$\pm$0 & 11.94 \\
\hline    
    \end{longtable}

\end{document}